\documentclass[useAMS,usenatbib]{mn2e}

\usepackage[T1]{fontenc}
 \usepackage{pslatex}
\usepackage{amsmath}
\usepackage{amsfonts}
\usepackage{xcolor}
\usepackage{url}
\usepackage{bm}
\usepackage{graphicx}
\usepackage{amssymb}
\usepackage{soul}
\usepackage{booktabs}
\usepackage{array}
\usepackage[utf8]{inputenc}
\inputencoding{latin1}
\inputencoding{utf8}

\newcommand{\fig}[1]{Fig.~\ref{fig:#1}}

\def\lsim{ \lower .75ex\hbox{$\sim$} \llap{\raise .27ex \hbox{$<$}} }
\def\gsim{ \lower .75ex \hbox{$\sim$} \llap{\raise .27ex \hbox{$>$}} }

\title[Magnetic reconnection and polarization in jets] 
{Kink-driven magnetic reconnection in relativistic jets: consequences for X-ray polarimetry of BL Lacs}

\author[Bodo, Tavecchio \& Sironi]
{G. Bodo$^1$\thanks{E--mail: gianluigi.bodo@inaf.it}, F. Tavecchio$^2$ and L. Sironi$^3$\\
$^1$INAF -- Osservatorio Astrofisico di Torino, Strada Osservatorio 20, I-10025 Pino Torinese, Italy\\
$^2$INAF -- Osservatorio Astronomico di Brera, via E. Bianchi 46, I--23807 Merate, Italy\\
$^3$Department of Astronomy and Columbia Astrophysics Laboratory, Columbia University, New York, NY 10027, USA\\
}

\begin{document}



\maketitle

\begin{abstract} 
We investigate with relativistic MHD simulations the dissipation physics of  BL Lac jets, by studying the synchrotron polarization signatures of particles accelerated by the kink instability in a magnetically-dominated plasma column. The nonlinear stage of the kink instability generates current sheets, where particles can be efficiently accelerated via magnetic reconnection. We identify current sheets as regions where $s=J\delta/B$ is above some predefined threshold (where $B$ is the field strength, $J$ the current density and $\delta$ the grid scale), and assume that the particle injection efficiency scales as $\propto J^2$. X-ray emitting particles have short cooling times, so they only probe the field geometry of their injection sites. In contrast, particles emitting in the optical band, which we follow self-consistently as they propagate away from their injection sites while cooling, sample a larger volume, and so they may be expected to produce different polarimetric signatures. We find that the degree of polarization is roughly the same between X-ray and optical bands, because even the optical-emitting particles do not travel far from the current sheet where they were injected, due to lack of sufficient kink-generated turbulence. The polarization angle shows a different temporal evolution between the two bands, due to the different regions probed by X-ray and optical emitting particles. In view of the upcoming {\it IXPE} satellite, our results can help constrain whether kink-induced reconnection (as opposed to shocks) can be the source of  multi-wavelength emission from BL Lacs. 
\end{abstract}


\section{Introduction}

According to the standard framework, relativistic extragalactic jets start in the vicinity of supermassive black holes as highly magnetized outflows, accelerating thanks to the conversion of Poynting flux into bulk kinetic flux under differential collimation. A direct prediction of this scheme is that the plasma remains magnetized, asymptotically reaching equipartition between the magnetic and the kinetic energy densities, until some dissipative processes occur \citep[e.g.][]{Komissarov07, Lyubarski10}.
The two main routes along which the jet power can be dissipated and channeled to energize ultra-relativistic particles (allowing the jets to shine from radio up to gamma rays) are collisionless shocks and magnetic reconnection, possibly triggered by global MHD instabilities (e.g. kink instability, \citealt{Begelman98}). 
However, shocks are thought to be rather inefficient for the large plasma magnetization expected in jets \citep[e.g.][]{Sironi15}. Moreover, while shocks convert only a fraction of the kinetic energy of the jet, reconnection is an efficient way to directly use the primary magnetic energy flux.  The idea that magnetic reconnection is a viable mechanism for energy dissipation and particle acceleration has received strong support by kinetic simulations showing that (relativistic) magnetic reconnection is able to energize a population of non-thermal particles, with an energy distribution following a power law \citep[e.g.][]{Sironi14, Guo14, Werner17, Petropoulou19}.

The best astrophysical sources in which our ideas on jet physics and particle acceleration can be tested are blazars \citep{UrryPadovani}. These peculiar active galactic nuclei are defined by the presence of a jet closely aligned toward the Earth. In this favorable geometry, the radiation from the jet is strongly beamed and often completely outshines the other components associated to the active nucleus. Blazars are further divided in Flat Spectrum radio Quasars (FSRQ) and BL Lacs. The latter subgroup contains the most extreme sources, copiously emitting in the very-high energy band above 100 GeV \citep[e.g.][]{Romero17}. Polarized emission is one of the defining properties of blazars \citep{AngelStockman80} and, as such, it has been extensively studied in the past decades. Ideally, polarimetric measurements can provide key information to characterize the magnetic field geometry associated to the innermost region of the flow. However, despite great efforts, the situation is still quite unclear. The regular and intensive monitoring, including polarimetric observations, triggered by the advent of {\it Fermi}-LAT led to the identification of possible regular patterns. In particular, the detection of systematic and large variations of the polarization angle, possibly in correspondence of large gamma-ray flares (e.g., \citealt{Blinov2018}), has been interpreted as the signature of an emission region moving along an helical path in the presence of a toroidal dominated magnetic field (e.g. \citealt{marscher2008,marscher2010,larionov2013}). Similar features are however also accounted for by a scenario in which most of the polarization variability is due to turbulence in the flow (e.g., \citealt{Kiehlmann2016,Kiehlmann2017}), possibly downstream a standing shock (\citealt{marscher2014,marscher2015}). In this framework the observed emission does not provide any indication on the structure of the magnetic field in the jet, since its properties are mainly shaped by the turbulent nature of the flow. 

So far, shocks have been considered the most plausible cause for energy dissipation and particle acceleration in blazar jets \citep[e.g.][]{Blandford19}. This is supported by the observed ejection of new radio-knots (interpreted as travelling shocks) at VLBI scale in correspondence to intense emission flares (e.g. \citealt{marscher2008,larionov2013}). Moreover, for BL Lac objects (especially for those emitting at high energy), the standard emission models point to rather low magnetic field intensities ($B\lesssim 0.5$ G are commonly inferred in TeV emitting BL Lacs, e.g \citealt{tavecchio2010}), compatible with shocks characterized by the required relatively large acceleration efficiency \citep[e.g.][]{Sironi15}. However, in considering these results it is important to remark that the physical quantities characterizing the emitting regions are often inferred by means of highly simplified, one-zone radiative models. Indeed, with more complex frameworks, assuming e.g. external radiation fields \citep[e.g.][]{TavecchioGhisellini2016} or anisotropies in the angular distribution of emitting particles \citep[e.g.][]{TavecchioSobacchi2020}, higher magnetic field are derived. Of particular relevance here are the recent results of \cite{Petropoulou19} and \cite{Christie19}, which demonstrate that the emission properties of BL Lacs can be satisfactorily reproduced by a scenario assuming magnetic reconnection and large jet magnetizations (with magnetic field intensities exceeding 1 G).

\citet{Tavecchio18} took a theoretically-inspired approach and considered how polarimetric studies of BL Lacs (in particular in the X-ray band) could help to test the two main scenarios for dissipation discussed above (i.e. shocks and instability-induced reconnection). In particular, they used a model inspired by particle-in-cell (PIC) simulations to investigate the polarimetric properties of the shock scenario. For the magnetic reconnection scenario, however, they were not able to perform a detailed study, concluding that  high-resolution, dedicated simulations are required (see also \citealt{Zhang17}). In this paper we explore the polarimetric signatures expected from magnetic reconnection triggered by kink instability, by means of high-resolution relativistic magnetohydrodynamic (RMHD) simulations.  As in \citet{Tavecchio18} we are particularly interested in applying the general results to BL Lac objects whose synchrotron emission peaks in the X-rays, quite interesting targets for the upcoming {\it IXPE} satellite \citep{weisskopf2016}, and to contrast them with those obtained in the case of the shock scenario.

The paper is organized as follows. In Section 2 we describe the simulation set-up used, in Section 3 we report the derived polarimetric behavior and in Sect. 4 we discuss our results.

\section{Simulations}
\label{sec:simul}

  \begin{figure}
  \includegraphics[width=\columnwidth]{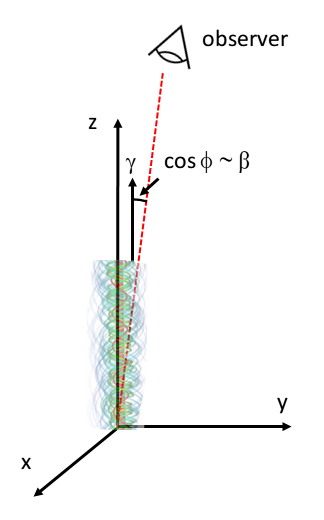} 
  \caption{Sketch of the geometry of the model. The jet moves along the $z$ direction with velocity $\beta$, corresponding to a Lorentz factor $\gamma = 10$. The observer lies in the $y-z$ plane and its line of sight forms an angle $\phi$ with the jet direction, such that $\cos \phi \sim \beta$.
  }
\label{fig:geometry}
  \end{figure}

\subsection{The Model}
\label{subsec:model}
Our aim is the study of the polarimetric signatures of the synchrotron radiation expected from reconnection triggered by the current driven kink instability. We then assume that a portion of  a relativistic jet, with an helical magnetic field, becomes subject to kink instabilities. The instability evolution promotes  the formation of reconnection regions, in which  the dissipation of magnetic energy causes the acceleration of relativistic particles \citep[e.g.][]{Sironi14, Guo14, Werner17, Petropoulou19} which then emit the high energy synchrotron radiation.   We study the evolution of the unstable jet portion by means of  three-dimensional  RMHD numerical simulations.  The jet moves with a bulk Lorentz factor of $\gamma \gg 1$, in the frame of the central black hole,  but the numerical simulations are performed in a reference  frame in which  the jet has a small velocity ($= 0.14 c$), in order to make contact with the linear analysis by \citet{Bodo13}.  We assume that the jet is magnetically dominated  and, although the simulation is performed with a non relativistic initial velocity, we have to use the relativistic form of the MHD equations since we will consider high values of the magnetization parameter. The processes that occur at the microscale in the dissipation regions cannot be captured by RMHD simulations;  PIC simulations, on the other hand, have shown that kink-induced reconnection regions have the ability to form a non-thermal population by accelerating particles to relativistic energies \citep{Alves18, Davelaar19}. At each time we  then inject in our RMHD simulation box,  in the regions that we identify as dissipation regions, a population of non-thermal particles, whose properties depend on the local conditions.   At high energies (X-rays emission) the lifetime of the emitting particles is much shorter than the dynamical time of the instability, the high energy part of the distribution function is then replaced at each time, depending on the local dissipation rate. At lower energies (optical emission) the radiative lifetime of the particles becomes comparable to the dynamical time, we have then to take into account the possibility that non-thermal particles  move  away from the injection region and accumulate over time. We can then compute the properties of the integrated emission as seen by an observer placed in the $y-z$ plane, where the direction $z$ is along the jet velocity. The line of sight of the observer forms an angle $\phi$ with the jet direction, such that $\cos \phi \sim \beta$,  so that in the simulation frame the line of sight form an angle of $\pi/2$ with the line of sight. A sketch of the geometry is presented in Fig.~\ref{fig:geometry}.

In the following subsections we will describe in detail all the steps outlined above for computing the polarimetric properties of the emitted radiation.

\subsection{Simulations set-up}

We model the unstable portion of the jet as a magnetized cylindrical region, with an helical magnetic field structure. The jet velocity is $v_j=0.14 c$ for $r < r_j$ ($r_j$ is the jet radius) and is connected to the outside medium at rest by a steep but smooth profile.   Consistently with the assumption of a magnetically dominated jet, we consider a force-free  initial configuration, described in detail in \citet{Bodo13}.   The  magnetic field has the following structure
\begin{equation} \label{eq:Bphi_prof}
 B_\varphi^2 = \frac{B_{\varphi c}^2}{(r/a)^2}\left[1 - \exp\left(-\frac{r^4}{a^4}\right)\right]\,,
\end{equation}
\begin{equation}\label{eq:Bz_prof}
  B^2_z = B^2_{\varphi c} \left[P_c^2 -  \frac{\sqrt{\pi}}{a^2}{\rm erf} \left(\frac{r^2}{a^2}\right) \right]
\end{equation}
\begin{equation}\label{eq:Br}
    B_r = 0
\end{equation}
where $z$ is the jet direction, $a$ is the magnetization radius, i.e. the radius inside which the magnetic field is concentrated, $B_{\varphi c}$ determines the maximum azimuthal field strength, $P_{c}$ is the value  of the pitch on the jet axis and $\mathrm{erf}$ is the error function. The magnetization radius $a$ is taken equal to $0.6 r_j$. The density $\rho_0$ and the pressure $p_0$ are uniform and since we consider a cold jet we take $p_0 = 0.01 \rho_0 c^2$. This equilibrium configuration is characterized by two parameters, $P_c$ and the  average jet magnetization $\sigma$. $P_c$ is defined as
\begin{equation}
    P_c \equiv \left| \frac{r B_z}{B_\varphi} \right|_{r=0} 
\end{equation}
while the average jet magnetization is
\begin{equation}
    \sigma = \frac{\langle B^2 \rangle}{\rho_0 h c^2}
\end{equation}
 where
\begin{equation}
    \langle B^2 \rangle = \frac{\int_0^a (B^2_z + B_\varphi^2) r dr}{\int_0^a r dr}
\end{equation}
and $h$ is the relativistic specific enthalpy. Note also that a factor $1/\sqrt{4 \pi}$ is absorbed in the definition of $B$.  The linear stability properties of such equilibrium configuration, generalized also with the inclusion of rotation, have been studied by \citet{Bodo13, Bodo16, Bodo19}. According to their results, in the parameter range considered here, the main instability is the current driven kink mode, while Kelvin-Helmholtz instabilities are unimportant.

For studying the evolution of the current driven kink instability we integrate the RMHD equations in the ideal limit (zero resistivity) by using the PLUTO code \citep{PLUTO}, adopting linear reconstruction, HLLD Riemann solver \citep{Mignone09} and the constrained transport method \citep{Balsara99, Londrillo04} to maintain a divergence-free magnetic field.   We make use of cartesian coordinates $x,y,z$, where $z$ is the jet direction. In the following we will measure lengths in units of $r_j$, time in units of $r_j/c$ and density in units of $\rho_0$.  The computational box covers a region $L \times L \times L_z$ with $480 \times 480 \times 800$ grid points, where $L=40$ and $L_z = 20$. The length of the domain has been chosen to be large enough to accomodate the mode with the fastest linear growth rate, which we found has also the fastest non-linear evolution. The grid is uniform in the region $|x| , |y| < 6$ (we have $x=y=0$ on the jet axis) and geometrically stretched elsewhere in order to reach larger transverse distances and avoid spurious effects from the lateral boundaries. The boundary conditions are periodic in the $z$ direction and outflows in all other directions. In order to visualize the jet deformation we also introduce a tracer, passively advected by the fluid and initially set to 1 inside the jet ($r < 1$) and to 0 outside. Ideal RMHD simulations are, in principle, scale invariant, however, for better clarity in the discussion, we can fix a scale of the simulation by choosing $r_j = 10^{16} \, \mathrm{cm}$, consequently the unit of time is $r_j/c = 3.33 \times 10^5 \mathrm{s}$.

\citet{Bodo13} have shown that the growth rate $\omega$ of the instability  scales as
\begin{equation}
    \omega \sim \frac{v_A}{P_c} \left( \frac{a}{P_c} \right)^2 f(k P_c)
\end{equation}
where $k$ is the longitudinal wavenumber of the mode and $v_A$ is the Alfv\'en velocity which, in the relativistic case is defined as function of $\sigma$ as
\begin{equation}
    v_A^2 = \frac{\sigma}{1 + \sigma}
\end{equation}
The function $f(k P_c)$ is independent from $P_c/a$ and it is a growing function of $k P_c$ up to $k P_c \sim 1$ where the growth rate drops to zero, the modes with a larger $k$ are then stable. The linear growth rate of the instability therefore increases as we increase $\sigma$ until it saturates for large values of $\sigma$ and also increases as we decrease $P_c$. Our equilibrium configuration has a minimum allowed value of $P_c/a$, namely for $P_c/a < \pi^{1/4}$ no equilibrium configuration is possible \citep{Bodo13}. 

Recently \citet{Bromberg19} have performed a series of simulations of the evolution of the current driven kink instability, comparing different initial equilibria.  They used  equilibrium solutions belonging both to the class of equilibria proposed by \citet{Mizuno09} and to the class of equilibria described above. The different equilibria present different pitch profiles and  \citet{Bromberg19}  pointed out the importance of the  profiles in determining the evolution of the instability, in particular the decreasing (with radius) pitch case appears to be the most efficient and fast in dissipating the magnetic energy. One of the differences between our equilibrium and the one proposed by \citet{Mizuno09} is the behavior of $B_z$ far from jet axis: in our case it tends to a constant value, while in Mizuno's case it tends to zero. However for $P_c/a = \pi^{1/4}$ the constant value is zero. In the present analysis we then choose a case with $P_c/a = 1.332$ very close to the minimum value, this ensures a very steep pitch profile,  the fastest linear growth,  and a very efficient energy dissipation. The other free parameter is $\sigma$, for which, consistently with our discussion, we choose a high value, namely $\sigma = 10$. We remind that this value of the magnetization is computed not with the peak values, but with the average value inside the jet. The use of the peak value would give a magnetization of $\sim 23$, while considering only the peak value of the azimuthal component of the magnetic field, which is the most likely component that gets dissipated,  we get $B_\varphi^2/\rho_0 h c^2 = 8.34$.

\subsection{The instability evolution}

\begin{figure*}
\hspace{-1truecm}
  \includegraphics[width=\textwidth]{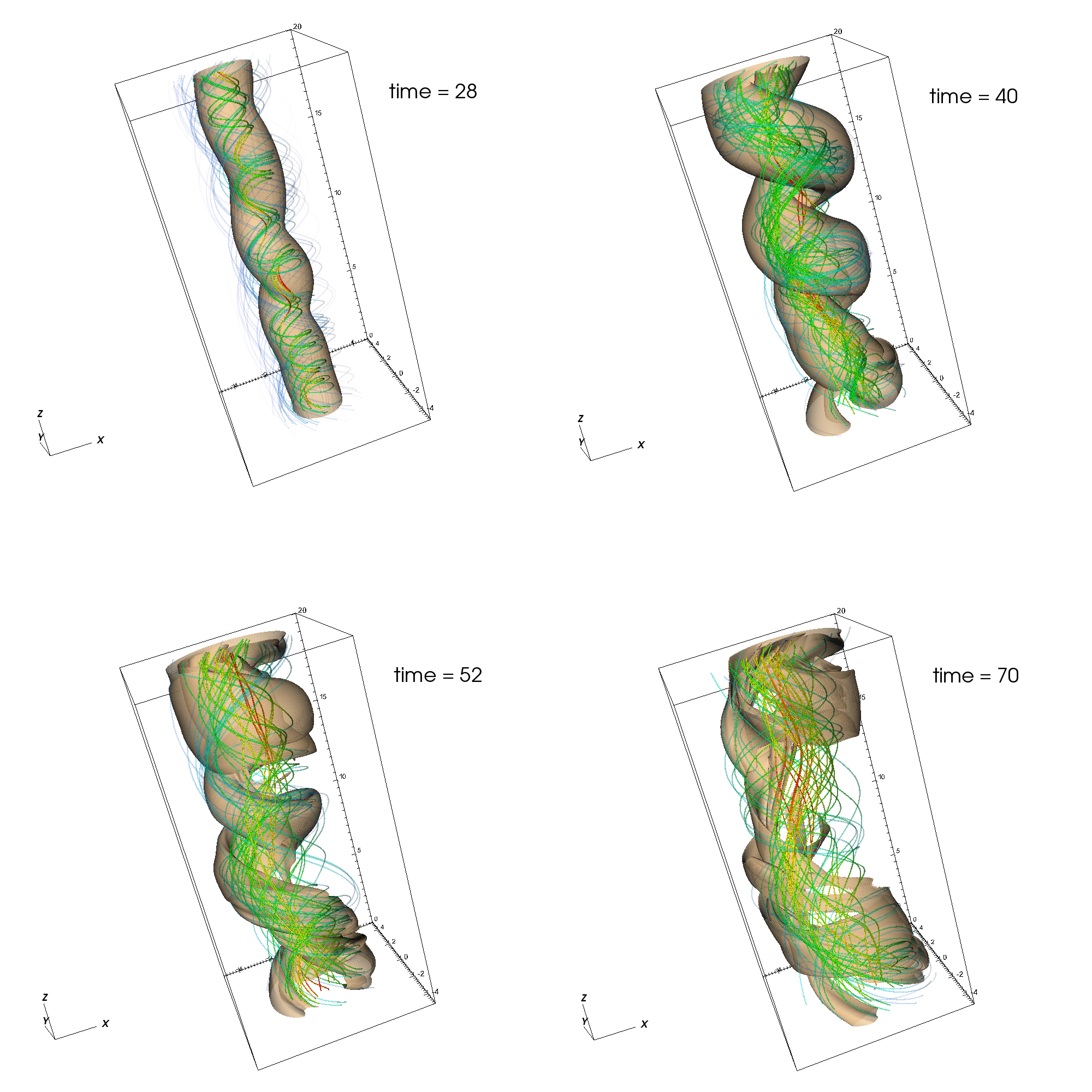} 
  \caption{The four panels represent different phases of the instability evolution. In each panel we show an isosurface of the tracer corresponding to a values of 0.9 and, superimposed, some representative magnetic field lines. The color of the field lines indicates the strength of the magnetic field. }
\label{fig:evol}
  \end{figure*}
  
  \begin{figure}
  \includegraphics[width=\columnwidth]{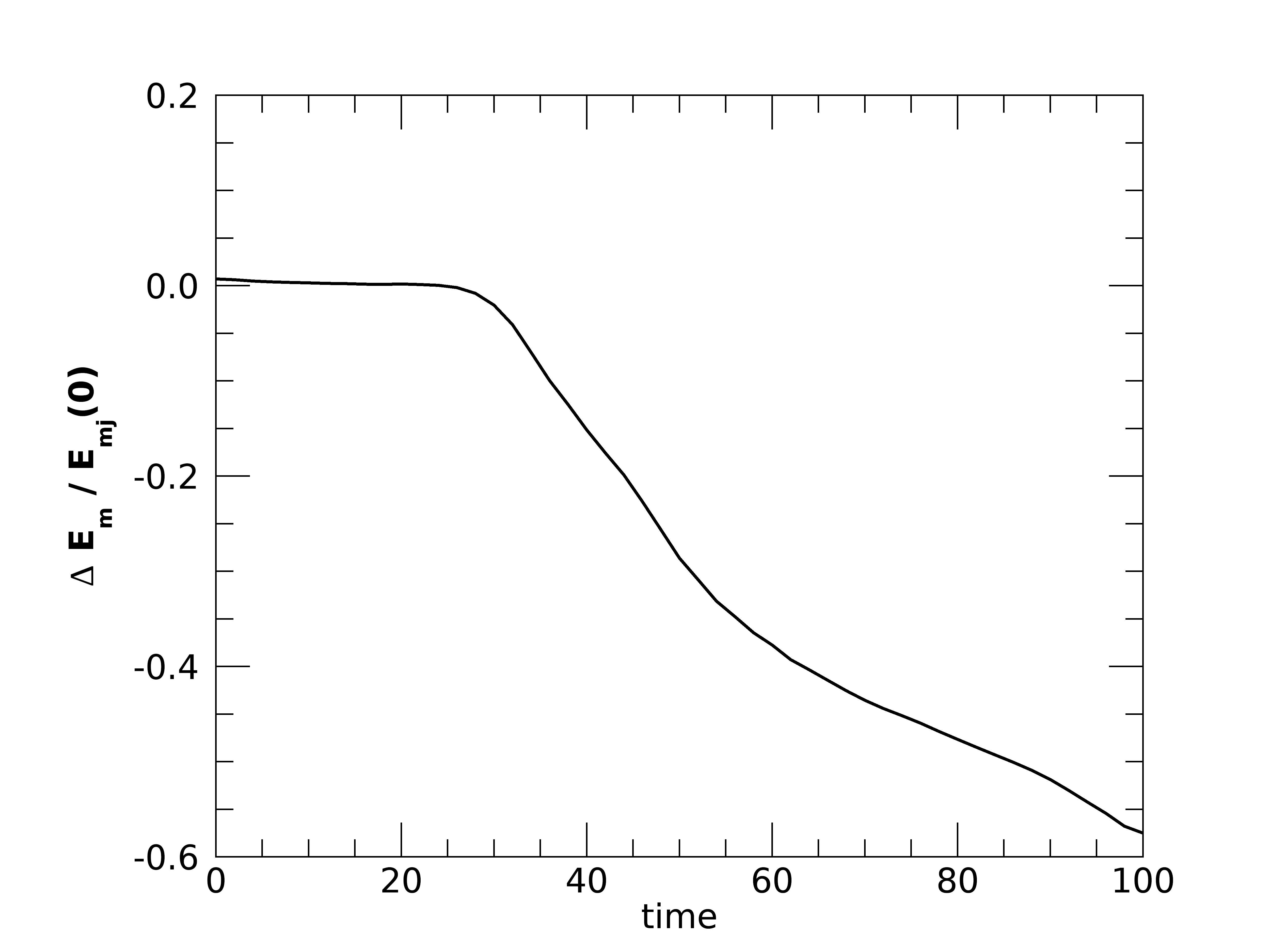} 
  \caption{Plot of the variation of the total electromagnetic energy as a function of time. The values are normalized to the electromagnetic energy in the jet at $t=0$ }
\label{fig:magen}
  \end{figure}

The equilibrium configuration described in the previous subsection is perturbed at $t=0$ by a random velocity perturbation. The random perturbation can be seen as the superposition of modes with different longitudinal wavenumbers $k$ and the choice of periodic boundary conditions in the $z$ direction limits the possible values of $k$ to the discrete set
\begin{equation}
    k = \frac{2 \pi n}{L_z}
\end{equation}
where $n$ is an integer number. From the results of the linear analysis, we have instability for $k P_c \lessapprox 1$, there will be then a set of unstable modes corresponding to all the values of $n$ up to a maximum and the growth rate increases with $n$.  For the case at hand,  this maximum is $n=4$. The modes with $n=3$ and $n=4$ have similar growth rate and the initial phase of the evolution of the instability will then be dominated by these two modes.  In Fig. \ref{fig:evol} we show the isosurface of the tracer corresponding to the value $0.9$, together with a sample of magnetic field lines, at four different times. In the top two panels we observe the growth of the helical jet deformation. In the top left panel, at $t=28$ we observe a superposition of the modes with the largest growth rate ($n=3$ and $n=4$). At $t=40$ (top right panel)  the mode with $n=3$ clearly dominates the evolution.  The growth of the helical deformations saturates at $t \sim 40$ and the system, as shown in the bottom two panels, evolves towards a quasi steady state, which, as discussed in detail by \citet{Bromberg19}, having an higher pitch, results to be stable to current driven instabilities, in the chosen computational domain. 

The evolution shown in Fig. \ref{fig:evol} is accompanied by a dissipation of the magnetic energy as shown in Fig. \ref{fig:magen}. In the figure we plot $(E_M - E_{M}(0))/E_{Mjet}(0)$ as a function of time, where $E_M$ is the electromagnetic energy integrated over the computational box and $E_{Mjet}(0)$  is the electromagnetic  energy in the jet at $t=0$. We note that, in the present situation, the dominant contribution to the electromagnetic energy comes from the magnetic component. We see that starting at $t \sim 30$ the magnetic energy has a steep decrease until $t\sim 60$, the decrease then proceeds at a slower rate. Most of the magnetic energy is dissipated and transformed into thermal energy,  a small fraction is converted into bulk kinetic energy, the resulting velocities are however subrelativistic with a maximum $\beta \sim 0.5$.  The first phase between $t\sim 30$ and $t\sim 60$ is the one in which strong current sheets are formed and almost $50\%$ of the initial jet magnetic energy gets dissipated. At later times dissipation proceeds for much longer times at a smaller rate and our current sheet detection algorithm (see below) does not find any current sheets.   The duration of the first phase, with strong current sheet, depends on the initial equilibrium configuration, for example \citet{Zhang17}, with different initial conditions find a somewhat longer duration. 

\subsection{The dissipation regions}
\label{sec:dissipation}

\begin{figure}
\includegraphics[width=\columnwidth]{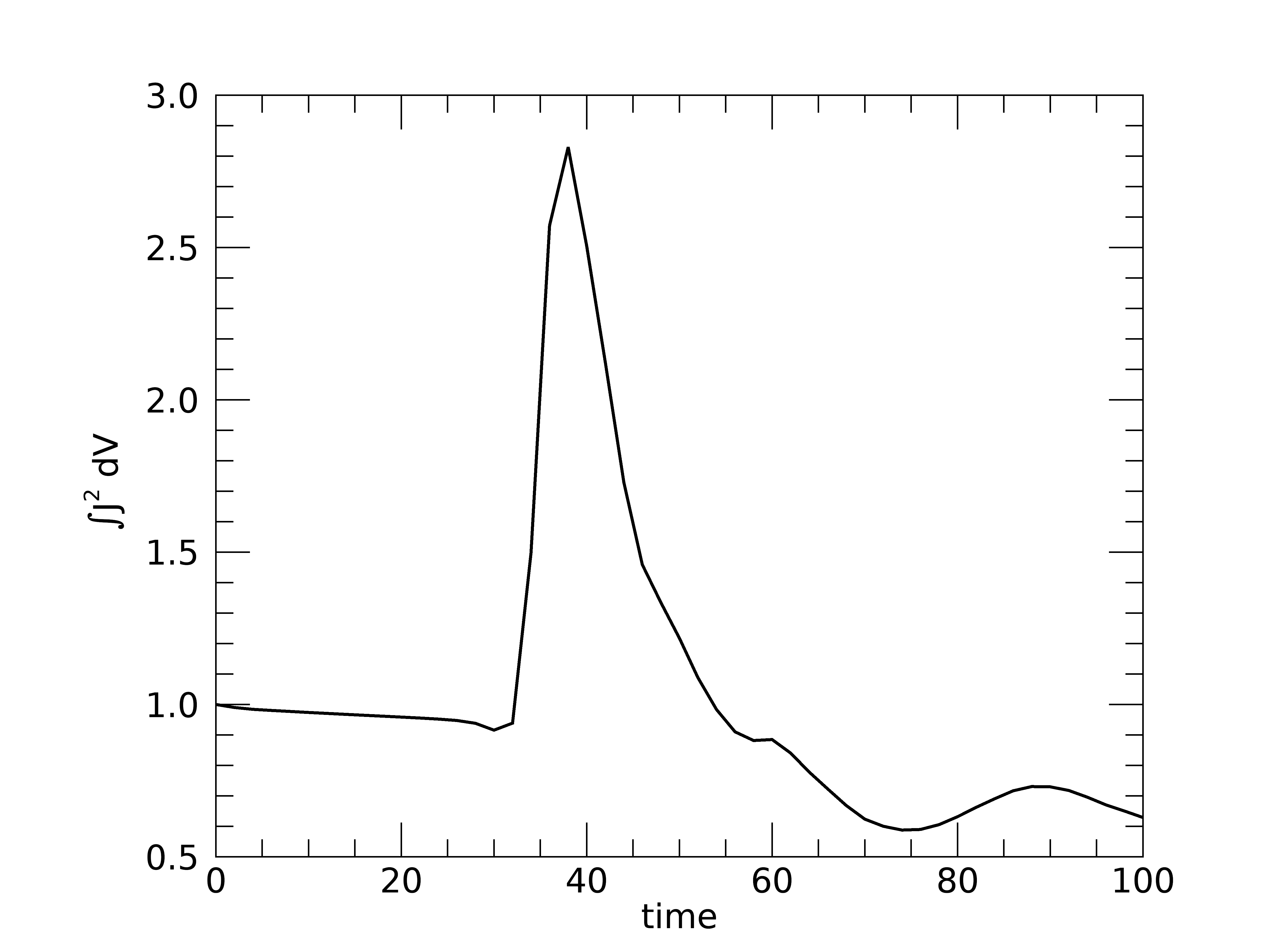} 
  \caption{Plot of the integral of $J^2$  over the computational box  as a function of time. The values are normalized to the value at $t=0$.}
\label{fig:j2}
  \end{figure}

Describing the instability evolution we have seen that its growth leads to a decrease of magnetic energy, which is partly dissipated and partly converted into bulk kinetic energy. For large magnetic Reynolds numbers dissipation occurs through the formation of current sheets, where magnetic reconnection takes place.  We notice that we solve the ideal RMHD equations with zero resistivity, so we rely on numerical dissipation for magnetic reconnection to occur. In the case of resistive simulations the energy density dissipation rate is given by $\eta J^2$, where $\eta$ is the magnetic diffusivity. Our simulations are not resistive and it is very difficult to quantify numerical dissipation, we can however take $J^2$ as a proxy for the dissipation rate \citep{Zhdankin15}. In fact comparing Fig. \ref{fig:magen} with Fig. \ref{fig:j2}, where we plot the integral of $J^2$ over the computational box as a function of time,  we can observe that the steepest part of the magnetic energy decrease corresponds to the peak of $\int J^2 dV$ between $t=30$ and $t=60$.

\begin{figure}
\includegraphics[width=\columnwidth]{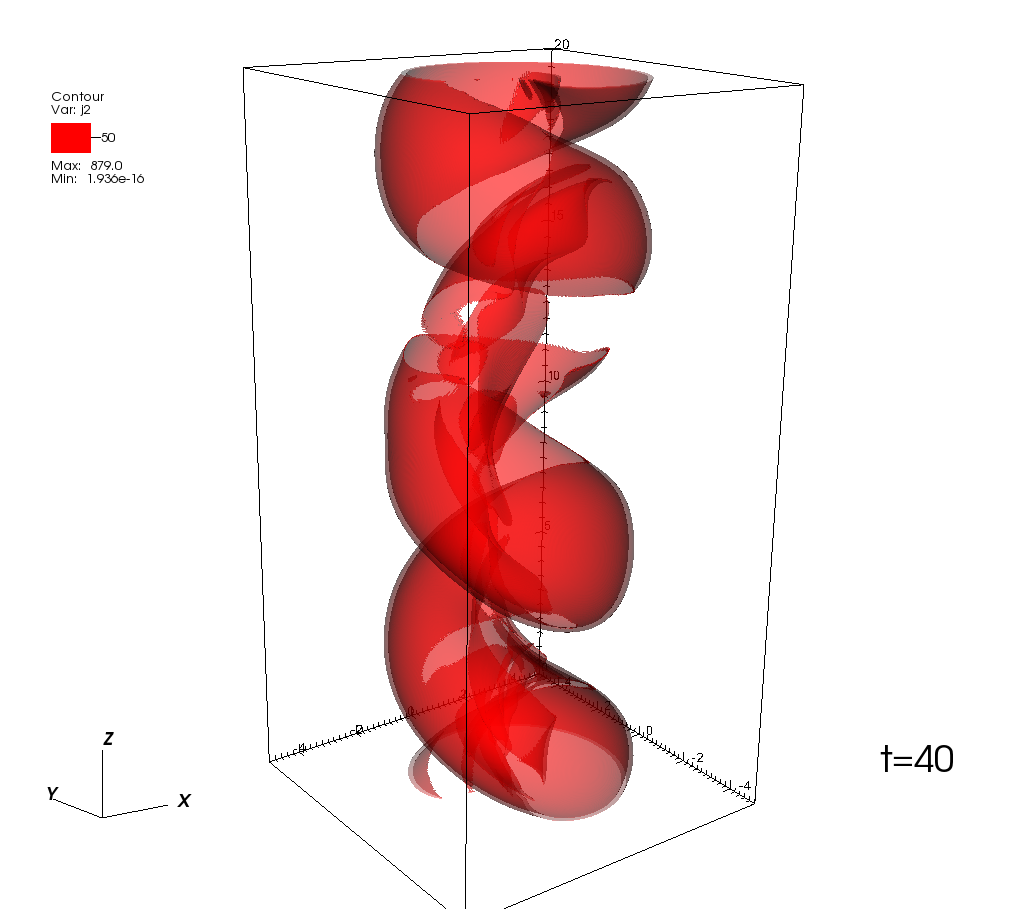} 
  \caption{The figure shows an isosurface of $J^2$ at $t=40$, around the time when the integral of $J^2$ reaches its maximum value. The isosurface represent the shape of the current sheets.}
\label{fig:isoj2}
  \end{figure}

\begin{figure*}
\vspace{-1.76truecm}
\hspace{-1truecm}
  \includegraphics[width=\textwidth]{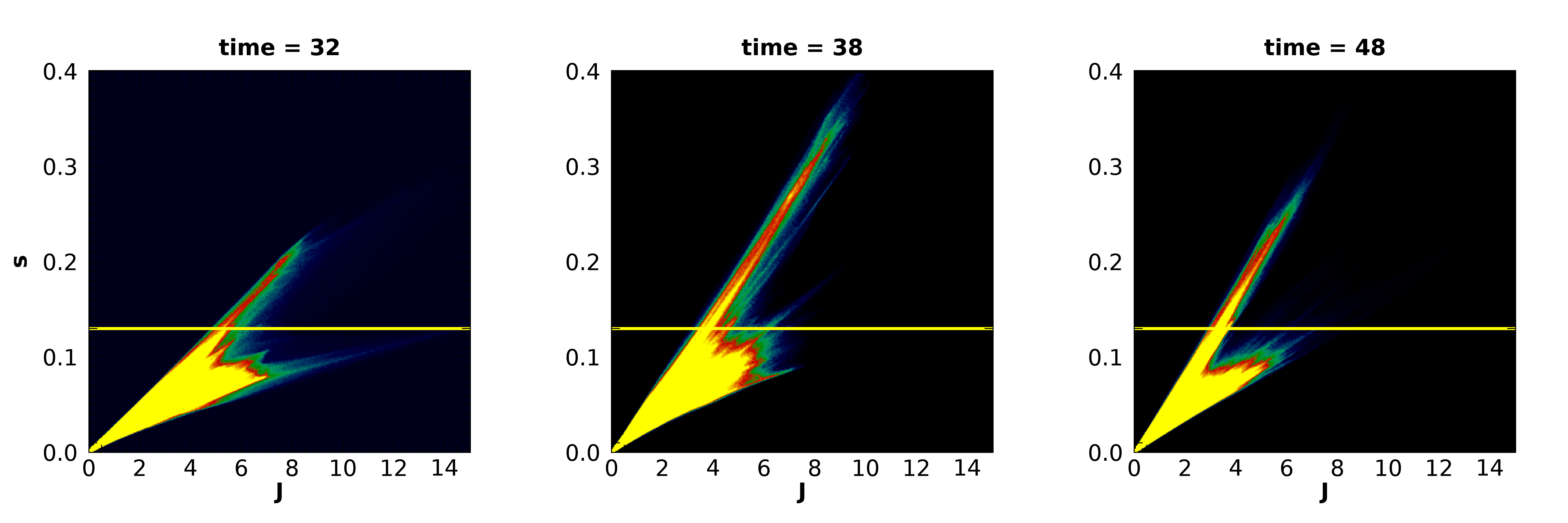} 
  \caption{Color map of the two-dimensional histogram of the points in the simulations, in the plane $J-s$. The three panels refer to three different times, from left to right $t=32$, before the maximum of the integral of $J^2$, $t=38$, around the maximum and $t=48$, in the decreasing phase. The horizontal line in each panel marks our threshold $s=0.12$.}
\label{fig:scale}  
  \end{figure*}

\begin{figure}
  \includegraphics[width=\columnwidth]{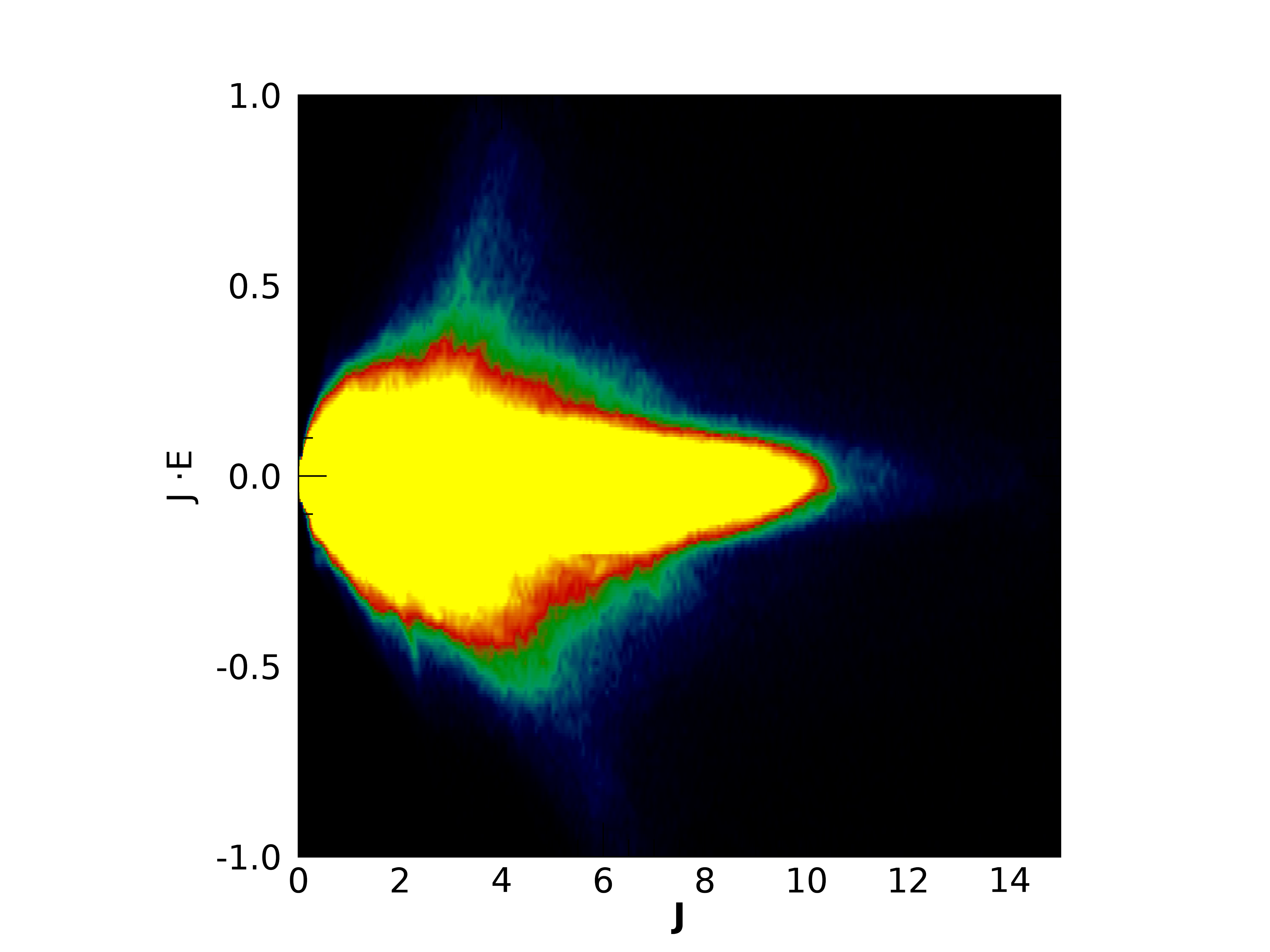} 
  \caption{Color map of the two-dimensional histogram of the points in the simulations in the plane $J-\mathbf{J}\cdot\mathbf{E}$ at $t=38$ around the maximum of the integral of $J^2$. }
\label{fig:je}
  \end{figure}
We can get a visual impression of the shape and locations of current sheets by displaying in Fig. \ref{fig:isoj2} an isosurface of $J^2$ at $t=40$ when the volume-integrated current reaches a maximum. Comparing Fig. \ref{fig:isoj2} with the top right panel of Fig. \ref{fig:evol} (both plots refer to the same time), we see that the current sheets form at the jet boundary and therefore acquire an helicoidal shape. For identifying and characterizing current sheets and dissipation regions one possibility would then be using the maximum values of the current \citep{Zhdankin15}. Alternatively,  \citet{Zhang17} quantify dissipation to be proportional to the quantity $\bf{J}\cdot \bf{E}$, where $\bf{E}$ is the electric field, which in an ideal simulation is $\bf{E} = -\bf{v}  \times \bf{B}$, and consider only the points where $\bf{J}\cdot \bf{E}$ has values above a given threshold. This choice, however,  identifies not only genuine dissipation, but also the work done by the magnetic field on the fluid.   We choose to follow a different approach. Taking into account that we rely on numerical dissipation, we identify as dissipative structures those defined on a small number of grid points, for which numerical dissipation becomes very effective. We then define a local steepness parameter as 
\begin{equation}\label{eq:scale}
    s = \frac{J \delta}{B}
\end{equation}
where $\delta$ is the cell size. $s$ represents a measure of the steepness of magnetic field gradients, the larger is $s$ the smaller is the number of grid points that locally resolve the magnetic field gradient.  We then choose to identify cells belonging to a current sheet as those that have $s$ larger than some threshold. We choose a threshold value of $0.13$, which corresponds to a resolution of about $8$ points, where numerical dissipation can be quite efficient \citep{Rembiasz17}. In order to verify the validity of our criterion and how it compares to the one based on the value of the current, in  Fig. \ref{fig:scale} we display as a two-dimensional histogram the number of points with given values of $s$ and $J$. The three panels refer to three different times, at the beginning of the growth of the instability (left panel), at the maximum of dissipation (middle panel) and towards the end of the dissipation phase (right panel). The horizontal line in each panel marks our threshold on $s$: the points above the horizontal line are those identified as belonging to dissipative structures. We can observe that above our chosen threshold we have a well defined correlation between $J$ and $s$, thinner structures have higher values of the current, as expected for current sheets. Since, however, there is some arbitrariness involved in the choice of the threshold, we show, in the Appendix, the consequences of a variation in the threshold.  On the contrary, a criterion based on the value of $\bf{J} \cdot \bf{E}$, as the one used by \citet{Zhang17}, leads to a selection also of points with a low value of $J$, where dissipation is likely to be much less effective, as can be seen by looking at Fig. \ref{fig:je} where we display the distribution of points as function of   $\bf{J} \cdot \bf{E}$ and $J$, similarly to what we have done in Fig. \ref{fig:scale}.  \citet{Zhdankin15} proposed a detection algorithm for current sheets that first selects points of maximum of the current, then builds clusters of points around these maxima in which the current is above some predefined threshold.   We  compared our criterion with the outcomes of this algorithm  and we found that they give very similar results.
\begin{figure}
  \includegraphics[width=\columnwidth]{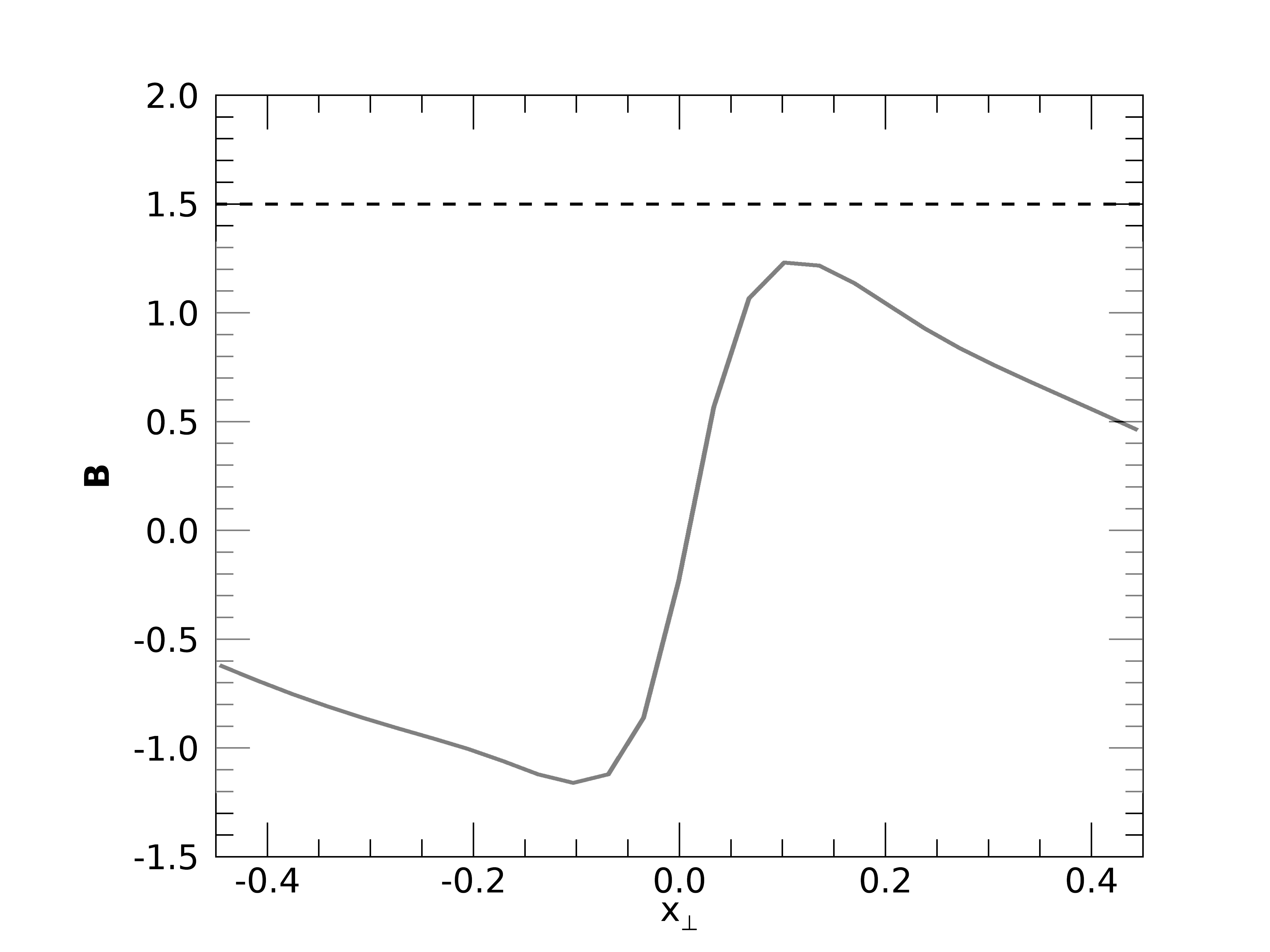} 
  \caption{The solid curve represents the variation of the reconnecting field in the direction normal to the current sheet at a selected position (the central point has coordinates $x=2$, $y=1$, $z=15$) at $t=40$, around the maximum of the integral of $J^2$. The dashed line indicates the value of the average guide field in the represented space interval. $x_\perp$ indicates the coordinate normal to the current sheet.}
\label{fig:prof}
  \end{figure}

In order to check that  the current sheets identified with our criterion effectively corresponds to regions where magnetic reconnection can take place, we have examined the profiles of the magnetic field in the directions orthogonal to the current sheets. In Fig. \ref{fig:prof} we show and example of such profiles, the solid line represents the magnetic field component that inverts its sign, while the dashed line represents the local guide field. As as side note, we point out that our procedure for identifying current sheets and for quantifying their role in particle energization implicitly accounts for the fact that current sheets with stronger guide fields are known to be less efficient in generating nonthermal particles \citep{Sironi15}. In fact, in our definition of $s$, we might estimate $J\sim B_{\rm rec}/\Delta$, where $B_{\rm rec}$ is the typical strength of the reconnecting magnetic field, and $\Delta$ the typical thickness of a current sheet. Within a current sheet, $B_{\rm rec}$ is small, and the magnitude of the magnetic field in the denominator of Eq.\ref{eq:scale} is then dominated by the guide field $B_g$. So, $s\sim (B_{\rm rec}/B_g) (\delta/\Delta)$. Assuming $\Delta\sim 3-4 \delta$, we see that $B_{\rm rec}/B_g\sim 0.5$ at our threshold value of $s=0.13$, and it linearly increases with $s$. Given the linear relation between $s$ and the electric current in Fig~\ref{fig:je}, and given that our energy injection rate of nonthermal particles is taken as $\propto J^2$, we are implicitly assuming that the efficiency of particle acceleration scales as $(B_{\rm rec}/B_g)^{-2}$.

\section{Polarization}
\label{sec:polariz}

    As discussed above, RMHD simulations cannot capture the microphysical processes that happen in the reconnection regions and in particular the formation of non-thermal  tails of the electron distribution function at relativistic energies. We then assume that at each time, at all points flagged as dissipation sites on the basis of the criterion discussed in the previous section,  a population of relativistic particles is injected. The form of the  energy distribution function is taken as $\propto K E^{-\alpha}$ \citep{Sironi14, Guo14, Werner17, Petropoulou19}. The normalization constant $K$ depends on the local energy dissipation rate, that we estimate as proportional to $J^2$.\footnote{Thus, we implicitly assume that particle acceleration primarily occurs at kink-generated current sheets, as recently suggested by PIC simulations (\citealt{Davelaar19}; see, however, \citealt{Alves18}, for alternative mechanisms of particle acceleration in kink-unstable jets, not associated with reconnecting current sheets).} As previously discussed, we assume that the portion of the jet subject to kink instabilities moves with a bulk Lorentz factor of $\gamma=10$. The relativistic particle population gives rise to synchrotron emission in the local magnetic field and this radiation is detected by an observer whose line of sight lies in the $yz$ plane and makes an angle $\phi$ with the jet velocity, with $\cos(\phi) = \beta$ (see also Fig. \ref{fig:geometry}). The polarization properties can be computed through the Stokes parameters obtained by integrating the emissivities over the computational domain (we checked that light-crossing time effects are negligible with the set-up we are considering). The Stokes parameters are  
    \begin{equation}\label{eq:IStokes}
        I(\nu) = \int \epsilon(\nu) dV
    \end{equation}
     \begin{equation}\label{eq:QStokes}
        Q(\nu) = \int \epsilon_{pol}(\nu) \cos(2 \psi) dV
    \end{equation}
         \begin{equation}\label{eq:UStokes}
        U(\nu) = \int \epsilon_{pol}(\nu) \sin(2 \psi) dV
    \end{equation}
    where $\epsilon(\nu)$ is the total, frequency dependent, synchrotron emissivity, $\epsilon_{pol}(\nu)$ is the polarized emissivity and $\psi$ is the local polarization angle that the electric field forms with the $x$ direction. The total and linearly polarized emissivities can be easily computed in the jet rest frame, where we perform the simulations, however they have to be transformed to the observer frame. Following \citet{Delzanna06} and \citet{Vaidya18} and, taking into account that, in the rest frame, the observer line of sight makes an angle of $\pi/2$ with the jet direction, the emissivities in the observer frame can be written as
    \begin{equation}\label{eq:em}
    \begin{aligned}
       & \epsilon(\nu) \propto \frac{4}{\alpha+1} \Gamma\left( \frac{\alpha}{4} + \frac{19}{12} \right) \Gamma\left( \frac{\alpha}{4} - \frac{1}{12}\right) \cdot \\
       & \cdot D^{2-(\alpha-1)/2} \, \nu^{-(\alpha-1)/2} J^2 \left( B_z^2 + B_x^2 \right)^{(\alpha+1)/4}  
    \end{aligned}
    \end{equation}
    \begin{equation}\label{eq:empol}
    \begin{aligned}
        &\epsilon_{pol}(\nu) \propto \Gamma\left( \frac{\alpha}{4} + \frac{7}{12} \right) \Gamma\left( \frac{\alpha}{4} - \frac{1}{12}\right) \cdot \\
        &\cdot D^{2-(\alpha-1)/2} \, \nu^{-(\alpha-1)/2} J^2 \left( B_z^2 + B_x^2 \right)^{(\alpha+1)/4}
        \end{aligned}
    \end{equation}
    where $\Gamma$ represents the $\Gamma$ function, $D$ is the Doppler factor,  $B_x$ and $B_z$ are the magnetic field components measured in the rest frame  and we assumed a power law distribution of the emitting electrons with normalization constant proportional to $J^2$.  For the transformation of the polarization angle  $\psi$ we follow \citet{Lyutikov03} \citep[see also][]{Delzanna06, Vaidya18} and we have that
    \begin{equation}\label{eq:cospsi}
        \cos(2 \psi) = \frac{B_z^2-B_x^2}{B_z^2+B_x^2}
    \end{equation}
    \begin{equation}\label{eq:sinpsi}
        \sin (2 \psi) = \frac{2 B_x B_z}{B_z^2+B_x^2}
    \end{equation}
    where the magnetic field components are measured in the simulation frame. We note that in the Lorentz transformations of the magnetic field components, leading to Eqs. \ref{eq:cospsi} and \ref{eq:sinpsi} we neglected the electric field observed in the simulation frame. This approximation leads to an error lower than 5\%. 
    Inserting Eqs. \ref{eq:em}, \ref{eq:empol}, \ref{eq:cospsi} and \ref{eq:sinpsi} in Eqs. \ref{eq:IStokes}, \ref{eq:QStokes} and \ref{eq:UStokes} and performing the integration over the computational box we can derive the observed Stokes parameters $I$, $Q$ and $U$ from which we obtain the polarization fraction $\Pi$ and the polarization angle $\chi$ as
    \begin{equation}
        \Pi(\nu) = \frac{\sqrt{Q^2 + U^2}}{I}
    \end{equation}
    \begin{equation}
        \cos(2 \chi) = \frac{Q}{\sqrt{Q^2+U^2}}
    \end{equation}
    
    Notice that the dependence of the polarization fraction on the power law index $\alpha$ is relatively weak. Our results show that reconnection typically occurs in a configuration where the guide field is of the same order of the reconnecting field. In this case the results by \citet{Werner17} show in this case the electron power law has $\alpha \sim 3$ and then the results that we present below are for $\alpha = 3$.
    
\subsection{Emission properties}
In principle, the emission observed from the jet during the evolution of the instability involves three timescales, namely the instability timescale -- which directly determines the timescale regulating the injection of high-energy electrons, $t_{\rm inj}$ -- the light-crossing time, $r_j/c$, and the cooling time of the electrons (function of the electron energy), $t_{\rm cool}$. As shown above (see also Fig. \ref{fig:evol}), the development of the instability occurs on times corresponding to tens of light crossing times. On the other hand, as discussed in Section \ref{subsec:model} the electrons emitting in the X-ray band have a lifetime much shorter than the dynamical time, i.e for the X-ray emitting electrons we have $t_{\rm cool} \ll r_j/c \ll t_{\rm inj}$. As a quantitative example, consider electrons emitting at an observed energy of 1 keV in a magnetic field of $B=2.5$ G, as adopted by \cite{Christie19} in their magnetic reconnection scenario for BL Lacs. The cooling time derived with the standard synchrotron theory is $t_{\rm cool}\simeq 3 \times 10^3$ s (where we assume that the emitted frequency is boosted by a Doppler factor $D=10$ and that inverse Compton losses can be neglected, since for electrons of these energies scatterings likely occur deeply into the Klein-Nishina regime).
Since the radius of the jet commonly derived for these sources is of the order of $r_j\approx 10^{16}$ cm we have, as assumed, $t_{\rm cool}\lesssim r_j/c$. 

\begin{figure*}
\vspace{-1.76truecm}
  \includegraphics[width=\columnwidth]{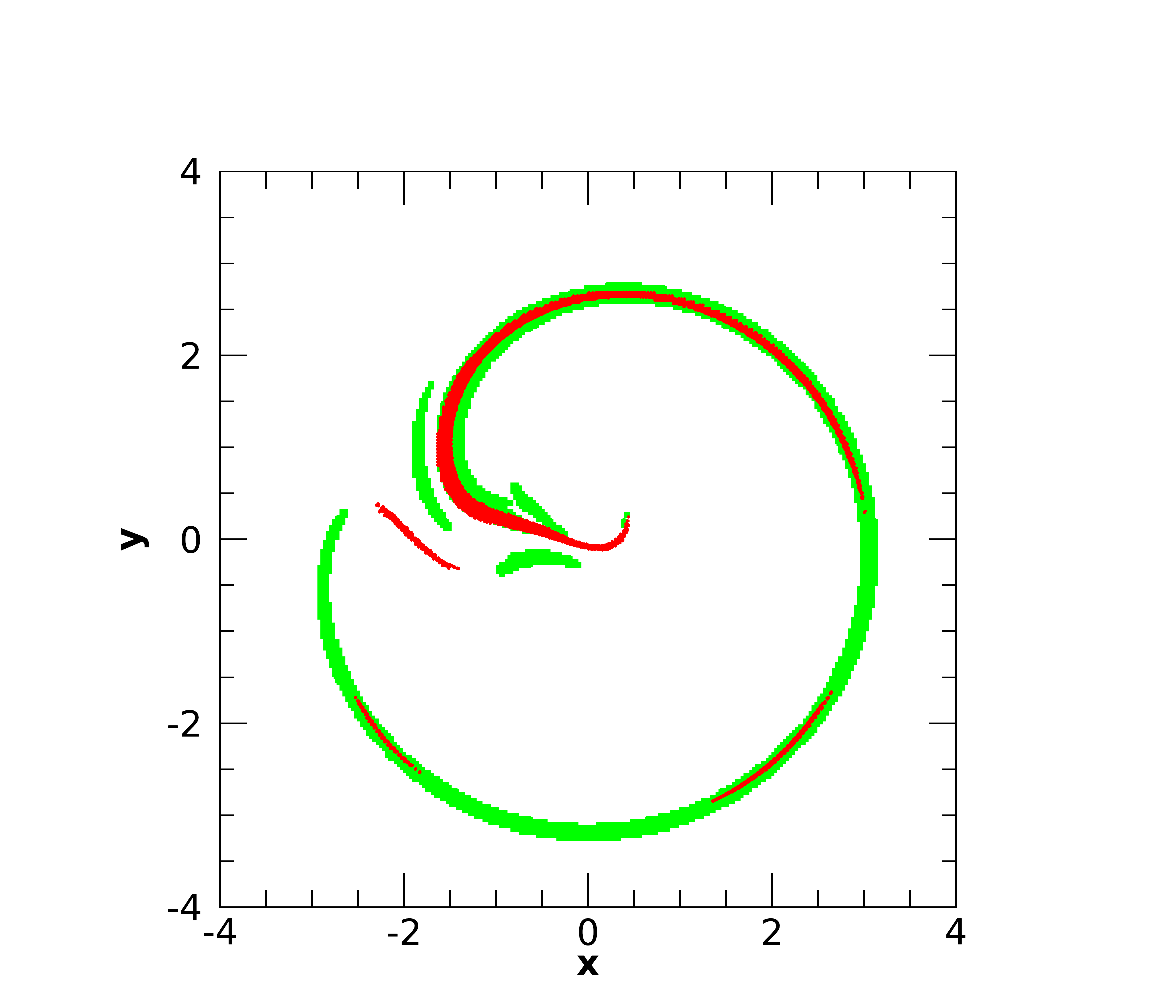} 
   \includegraphics[width=\columnwidth]{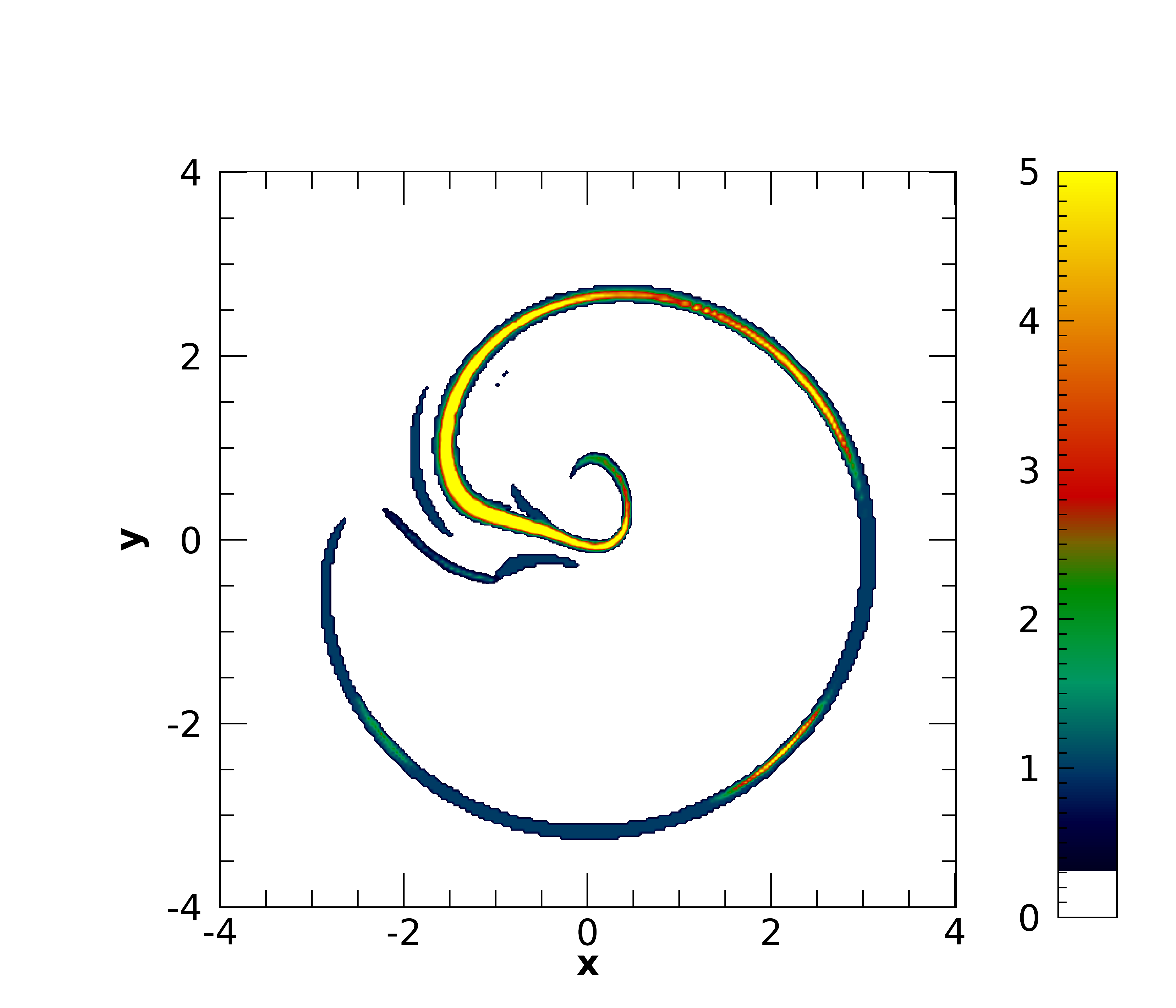} 
  \includegraphics[width=\columnwidth]{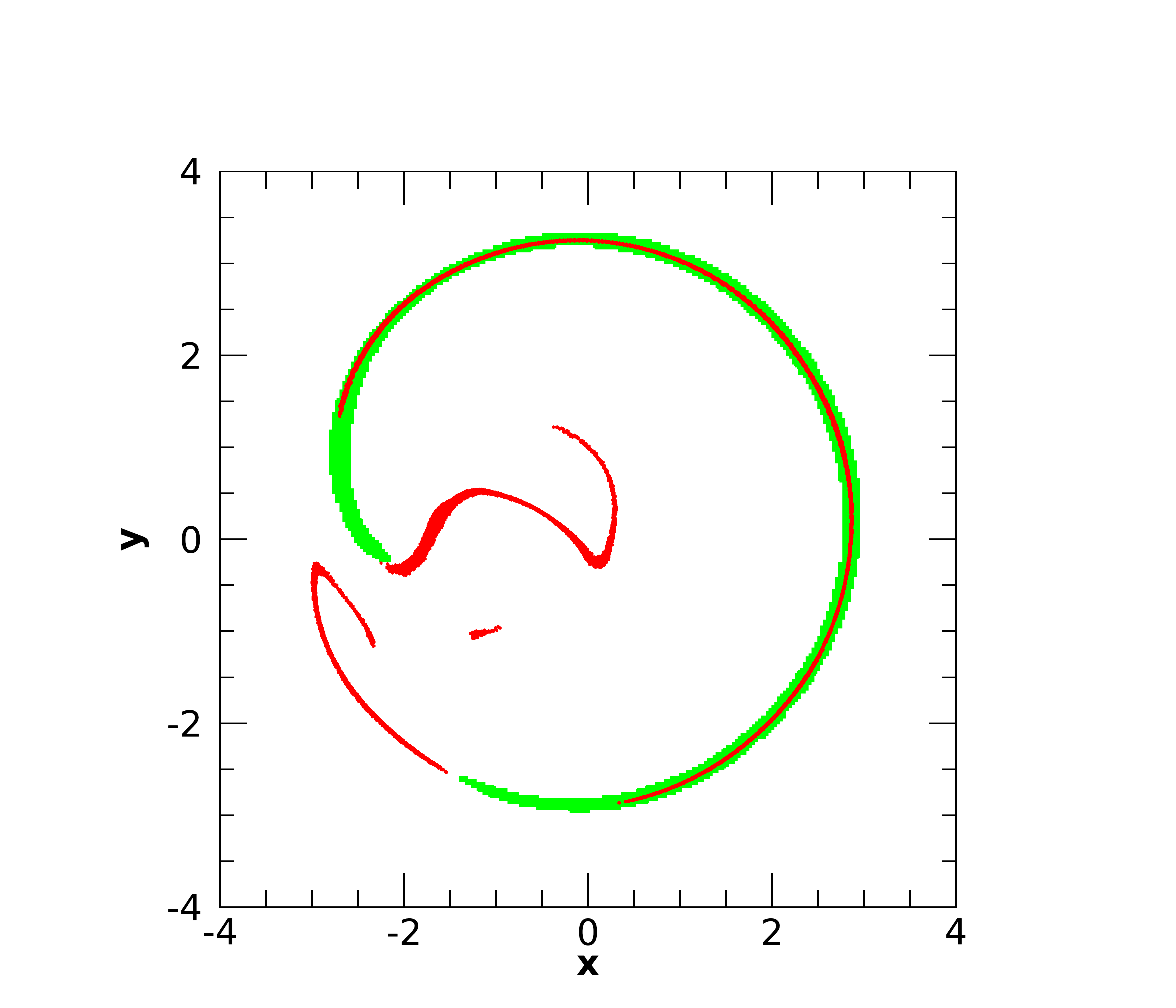} 
   \includegraphics[width=\columnwidth]{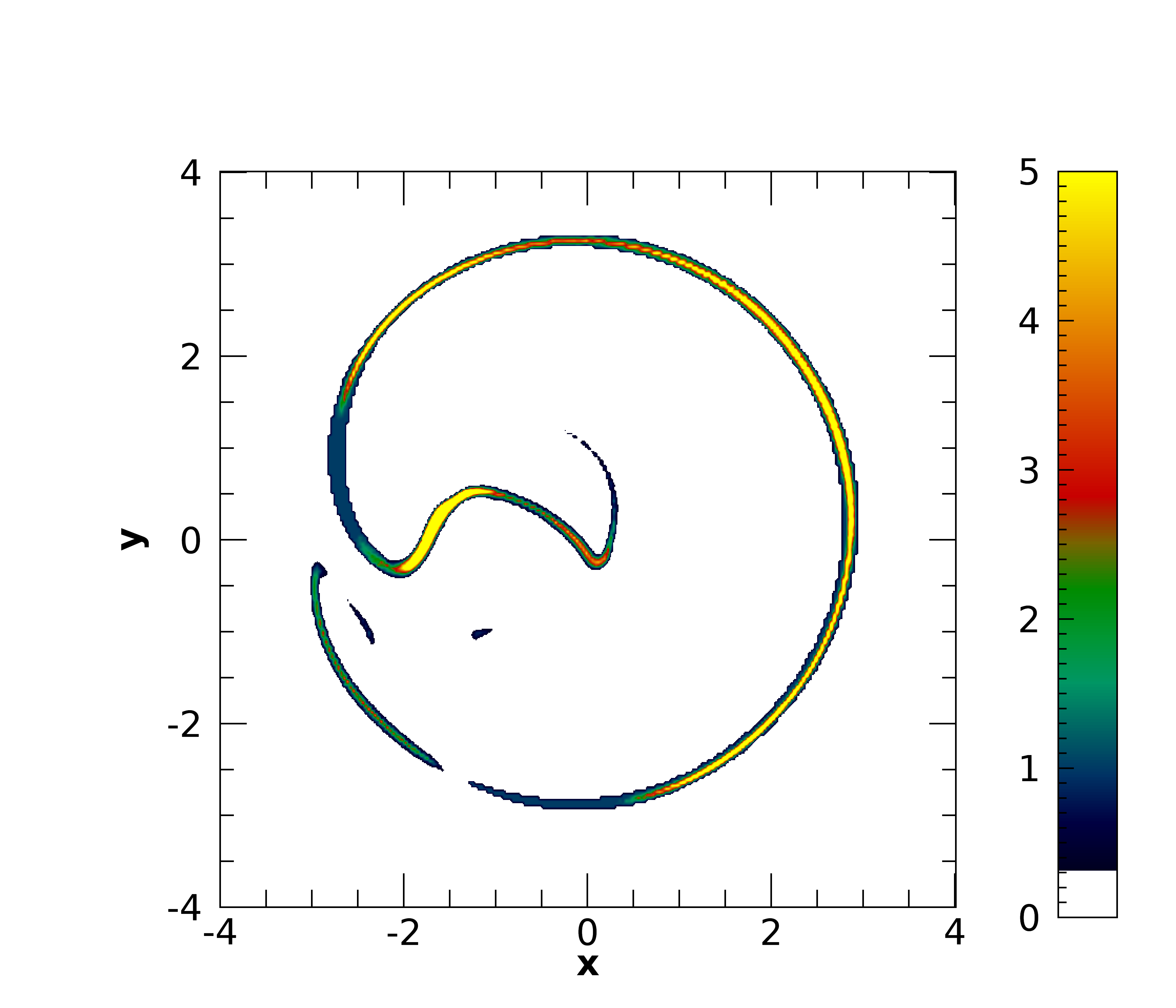} 
 
  \caption{The panels on the left show the positions of the particles in a transverse ($x-y$) section of the jet at $z=10$, in green we show particles that have an age less than $r_j/c$ and in red the particles that have an age between $3 r_j/c$ and $10 r_j/c$. The panels on the right show instead the density of particles in the same section of the jet. The panels on top refer to $t=40$, around the maximum of the integral of $J^2$, while the panels at the bottom refer to $t=46$, in the decreasing phase of the current. }
\label{fig:particles}
  \end{figure*}

The condition  $t_{\rm cool}\lesssim r_j/c$ allows us to compute the radiative properties of the X-ray emission at each time by considering just the particles freshly injected by the acceleration process in the dissipation regions, and, as discussed above, we assume the injection rate to be proportional to $J^2$ that is taken as a proxy for the local energy dissipation rate. In contrast, emission at lower energy involves particles that have longer lifetimes and therefore can propagate away from the regions where they are generated. For instance, with the parameters used above, electrons radiating in the optical band have a cooling time of the order of $2\times 10^6$ s, i.e. $t_{\rm cool}\approx 10 r_j/c$, i.e. of the order of the evolution timescale. In order to study the spatial distribution of these particles we made the assumption that they are carried by the fluid, i.e. that they move on average at the same velocity as the fluid. This assumption is well justified. For the times considered in this paper, the development of the kink mode has not resulted in volume-filling turbulence. If strong turbulence were to be generated, the turbulent power cascading down to scales comparable to the Larmor radius of the emitting electrons might enhance their diffusive transport. In the absence of significant kink-induced turbulence, as we observe, and assuming for simplicity that the accelerated particles do not self-generate appreciable turbulence (e.g., out of anisotropies in their distribution function), the particles will likely move together with the MHD fluid. We therefore introduced lagrangian markers that are injected in the dissipation regions identified following the prescription described in subsection \ref{sec:dissipation} and then move with the fluid velocity. Each lagrangian marker carries the time of injection $t_{inj}$ and the value of $J^2$ at $t_{inj}$, from these values we can then reconstruct the properties of the synchrotron emission in the optical band.
At a given time $t$ we can then obtain the distribution of particles of a selected age $\tau = t-t_{inj}$. An example of such distributions is shown in Fig. \ref{fig:particles}, where, in the left panels, we show the particle distributions in a transverse ($x-y$) cut of the jet at $z=10$ at two different times, at $t=40$ (upper panel), which is around the maximum of the dissipation, and at $t=46$ (lower panel), which is in the decreasing phase. In the figures we show in green the positions of  particles with age $\tau < r_j/c$  and in red the positions of particles with $3 r_j/c < \tau < 10 r_j/c$ (green). In the right panels we show instead the density of particles for the same sections and same times.  Young particles trace the strongest current sheets, while older particles tend to accumulate in some areas particularly at the two opposite edges of the current sheet, but they do not move away from the sheet in the transverse direction. Comparing the panels on the left with those on the right, we can see that the regions where the particles tend to accumulate (higher polarization density, yellow regions in the right panels) corresponds, as expected, to the regions where we find older particles (red regions in the left panels).

\begin{figure}
  \includegraphics[width=\columnwidth]{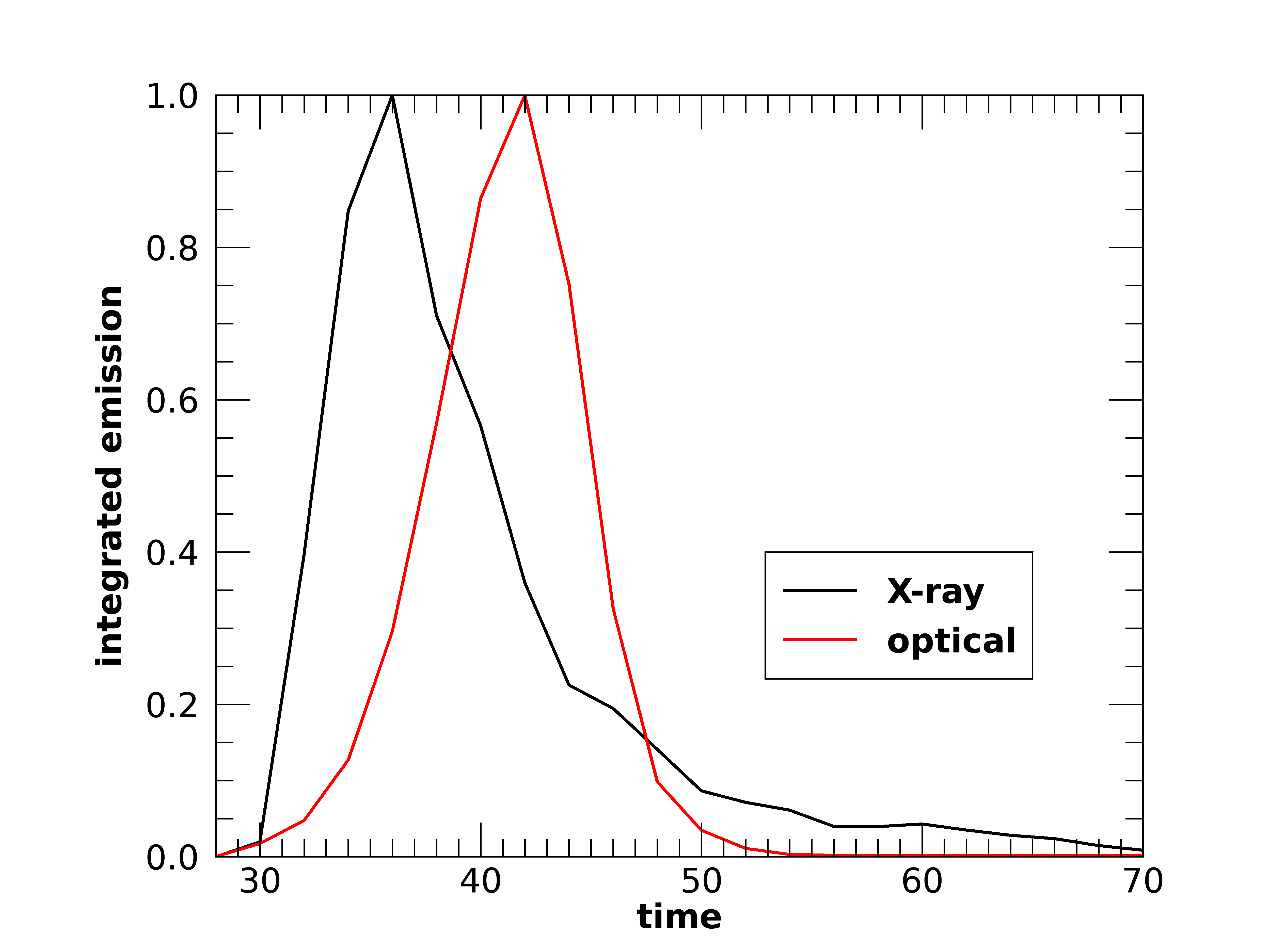} 
  \caption{Plot of the total emission in the X-ray (black) and optical (red) bands as a function of time. Each curve is normalized to its maximum value.}
\label{fig:flux}
  \end{figure}

A last important consideration
that simplifies the calculations is that, as mentioned above, the timescale of the instability (and therefore the injection timescale) is much larger than both the light-crossing time and the cooling time of the X-ray electrons. In particular, the conditions $t_{\rm cool}\lesssim t_{\rm inj}$, and $r_j/c\lesssim t_{\rm inj}$ imply that the modulation of the X-ray emission follows the evolution of the instability. This can be clearly appreciated in Fig. \ref{fig:flux}, which shows an extended emission tail in the X-rays at late times -- with a duration much longer than the lifetime of the high-energy electrons -- clearly following the decay of the dissipated energy (Fig.\ref{fig:j2}). The key point is that since the X-ray emission is modulated over timescales of the order of the instability times we are allowed to neglect the light crossing time on the observed properties of the emitted X-rays. Indeed, the inclusion of the light crossing time effects would only introduce a general smoothing of the lightcurve but it would not change the gross features that we obtain.
For the emission in the optical band the situation is different, since, as detailed above, in this case the cooling time is larger than the light crossing time. This implies that the properties of the optical emission are mainly determined by the competition between the cooling and the injection timescales and therefore light crossing time effect can be safely neglected. 
  
\begin{figure}
  \includegraphics[width=\columnwidth]{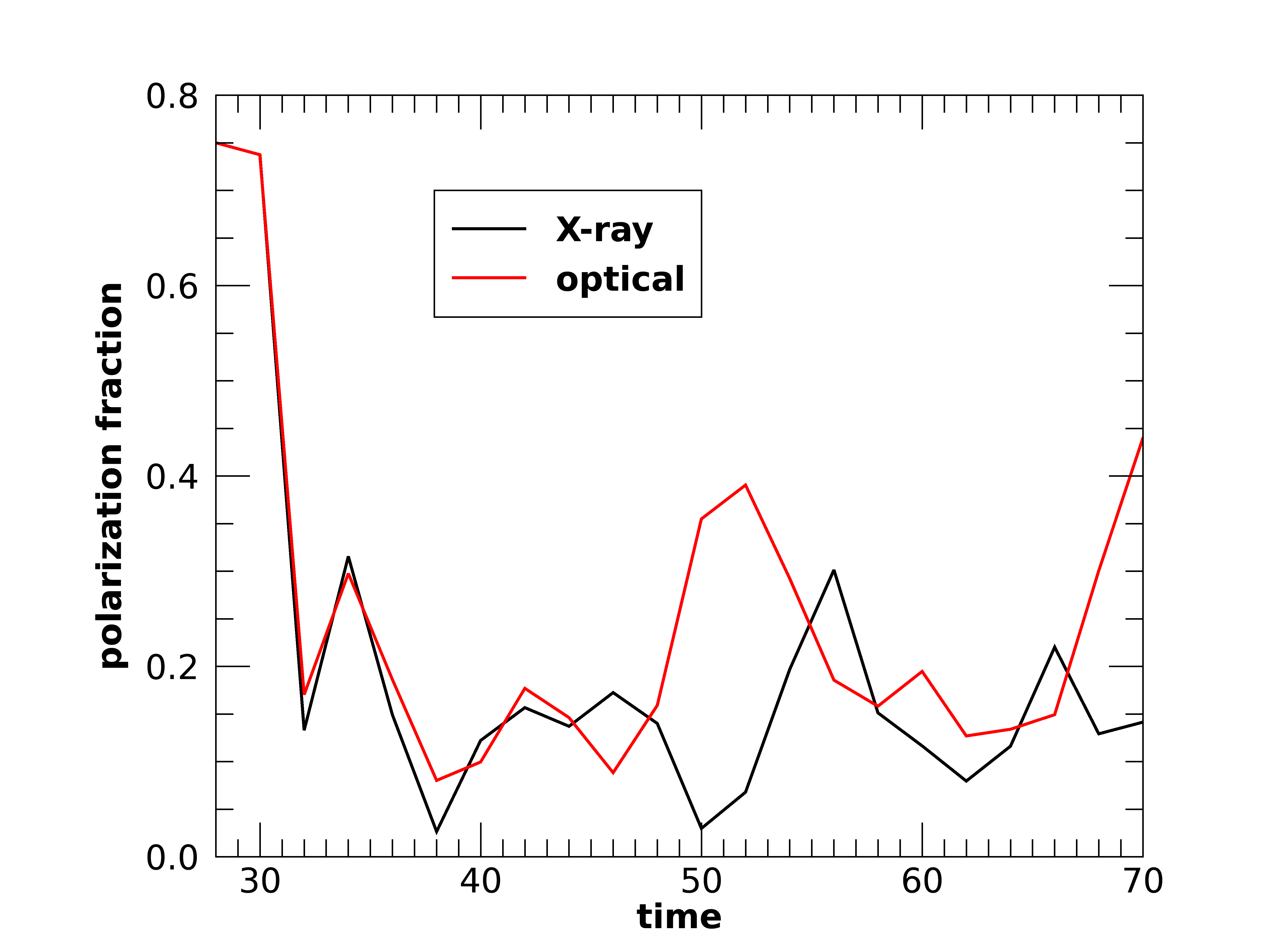} 
  \caption{Plot of the polarization fraction in the the X-ray (black) and optical (red) bands as a function of time.}
\label{fig:polfrac}
  \end{figure}

In Fig. \ref{fig:flux} we plot the Stokes parameter $I$, i.e. the total emission, as a function of time: the black curve refers to  X-rays, while the red curve refers to the optical band, both normalized to their maximum values.  The  X-ray emission has a sudden rise at $t=30$, peaks at $t=36$ and then declines more slowly up to $t=70$, after which it becomes negligible. The X-rays give a snapshot of the instantaneous dissipation rate, while the emission in the optical represents an integral of the dissipation rate over a longer time ($\sim 10$). During this interval the emitting particles accumulate, and therefore the optical lightcurve rises more slowly and the peak occurs at a later time.  In Fig. \ref{fig:polfrac}, which represents the polarization fraction as a function of time, we observe that, at the beginning of the flare, the polarization fraction decreases as the flux increases, then it varies in an erratic way, with values ranging from less than $5\%$ to about $40\%$. The polarization fraction in the optical follows almost exactly the behaviour of the X-rays up to $t=45$; afterwards, the two curves follow different paths, with a maximum of polarization in the optical corresponding to a minimum in the X-rays at $t \sim 50$.

\begin{figure*}
\vspace{-1.76truecm}
\hspace{-1truecm}
  \includegraphics[width=\columnwidth]{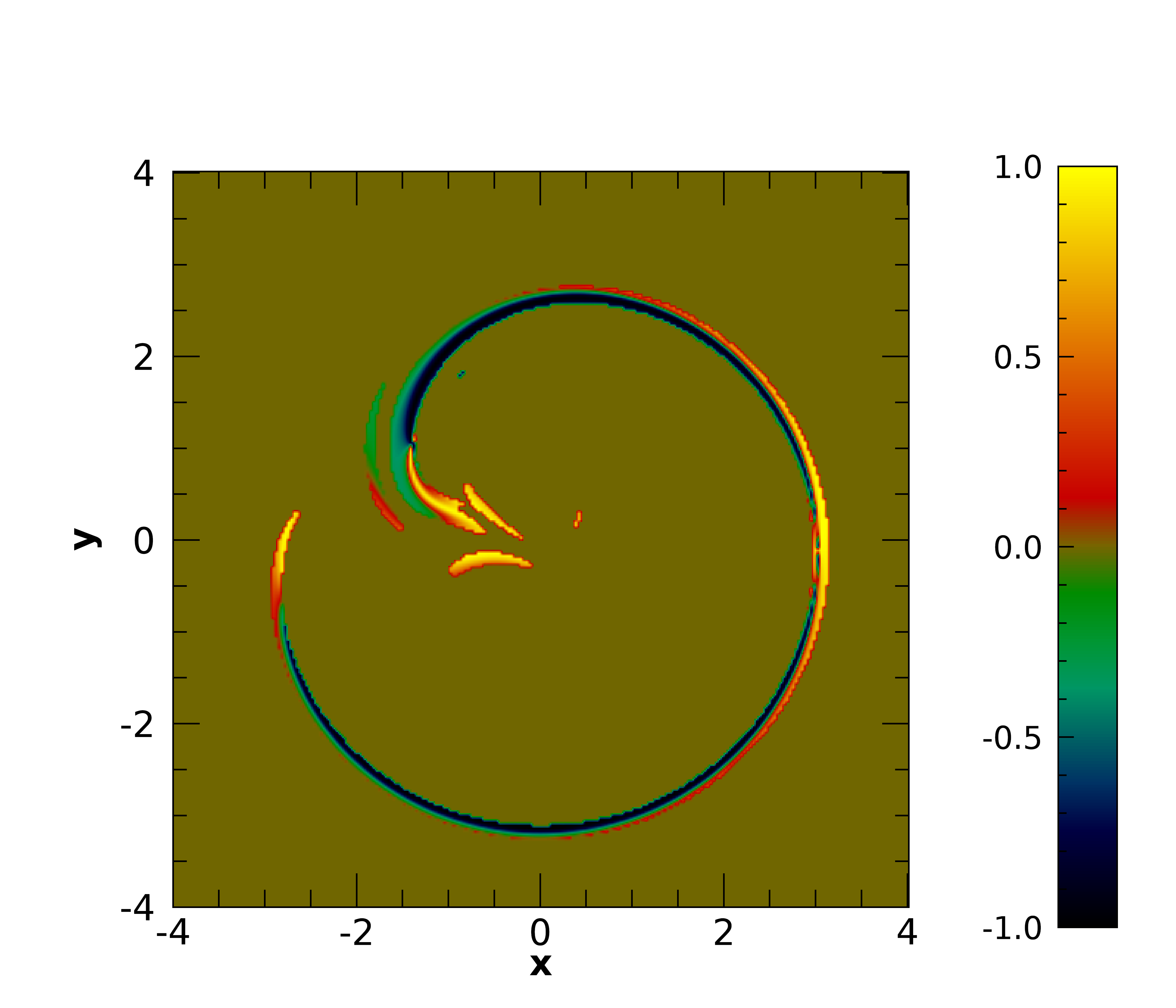} 
  \includegraphics[width=\columnwidth]{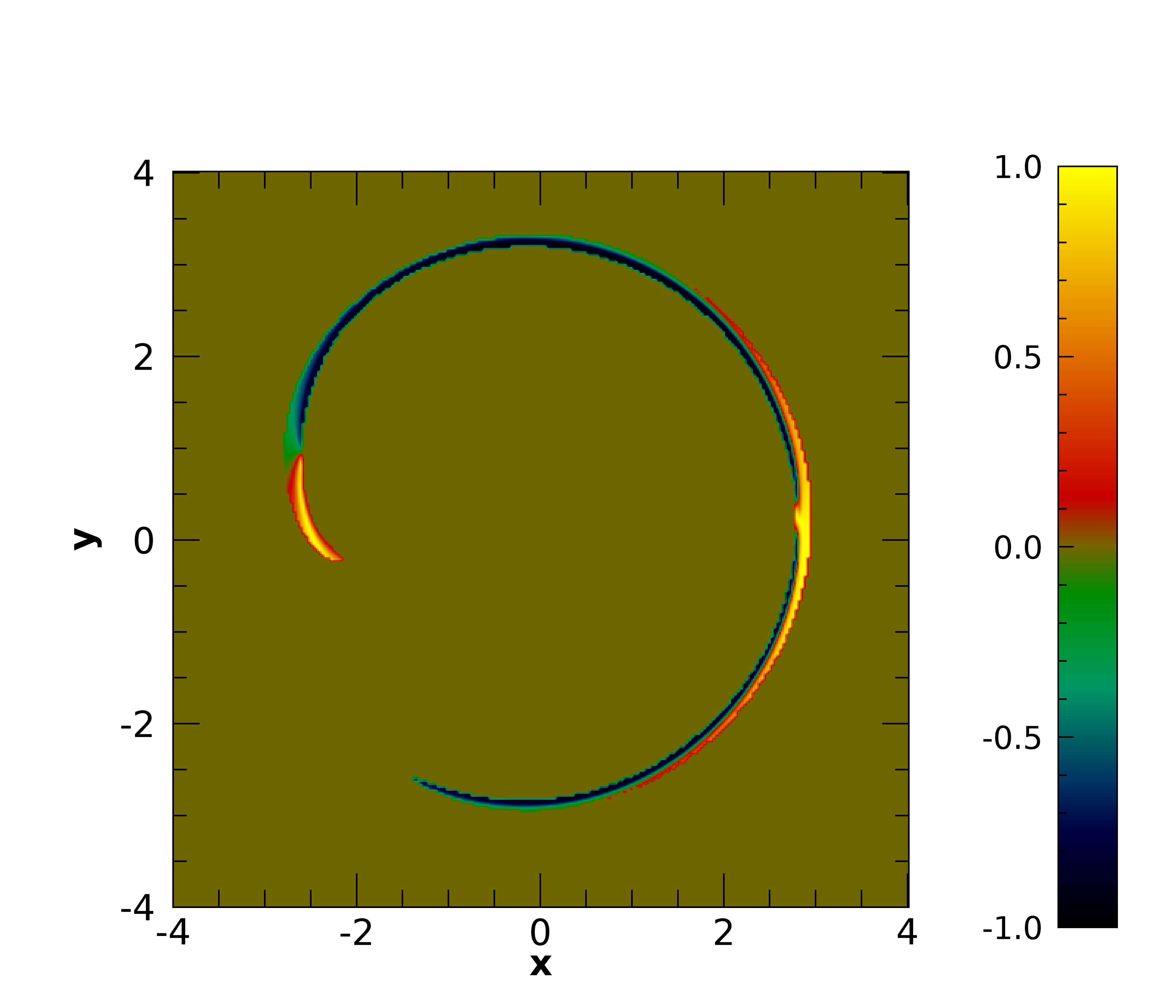} 
 
  \caption{Distributions of $\cos 2\psi$ in the current sheet on a transverse ($x-y$) section of the jet at $z=10$. The two panels refer to two different times, $t=40$ (left panel), around the maximum of the integral of $J^2$, and $t=48$ (right panel), in the decreasing phase of the current. The angle $\psi$ is the angle formed by the electric field with the $x$ axis, therefore the value $\cos 2 \psi =1$ corresponds to a local polarization direction along the $x$ axis, while  $\cos 2 \psi =-1$ corresponds to a local polarization direction along the $z$ axis. 
  }
\label{fig:coschi} 
  \end{figure*}

 At the starting of the flare ($t\sim 30$) we have only a limited spatial region where reconnection occurs; the radiation that we observe depends only on the magnetic configuration in that region and therefore the polarization fraction is high. As the instability proceeds, an helicoidal current sheet forms all around the jet and the integrated emission samples different magnetic field orientations and the polarization fraction, which depends on the topology of magnetic field in the current sheet, drops.  In Fig. \ref{fig:coschi} we show a transverse cut of the distribution of $\cos 2 \psi$ (see Eqs. \ref{eq:cospsi}) in the current sheet, as detected by our criterion (see Eq. \ref{eq:scale}). We focus on the Stokes parameter $Q$ since its contribution to the polarization fraction is the dominant one. We recall that we assume the line of sight to be along the $y$ direction and the angle $\psi$, as defined in Eq. \ref{eq:cospsi}, is such that $\cos 2\psi \sim 1$ when the dominant component of magnetic field is $B_z$  while $\cos 2\psi \sim -1$ when the dominant one is $B_x$. 

The two panels in the figure represent cuts taken at $z=10$ at two different times ($t=40$ around the peak of emission and $t=46$ during the decreasing phase of X-ray emission) and exemplify the distribution of the magnetic field direction. We can notice regions dominated by the $B_x$ component and regions dominated by the $B_z$ component. The contributions to the integrated emission by the two regions are almost equivalent and therefore the polarization fraction is low. As the balance between the contributions from the two different regions changes, we observe oscillations in the polarization fraction. Of course, there are also regions where the field has other orientations (green and red regions), they occupy a smaller area, but they determine the polarization angle (discussed below) when the contributions coming from yellow and blue regions ($\cos(2\chi) = \pm 1)$ balance. In the optical band, during the first part of the evolution, the distribution of the emitting particles follows quite closely the  particles emitting at higher energy and so does the polarization fraction. As the dissipation proceeds, the two distributions become different and sample different regions, and as a consequence the  behavior of the polarization fraction can be different with one or the other (i.e., X-ray or optical) being larger, depending on the sampled regions.  

\begin{figure}
  \includegraphics[width=\columnwidth]{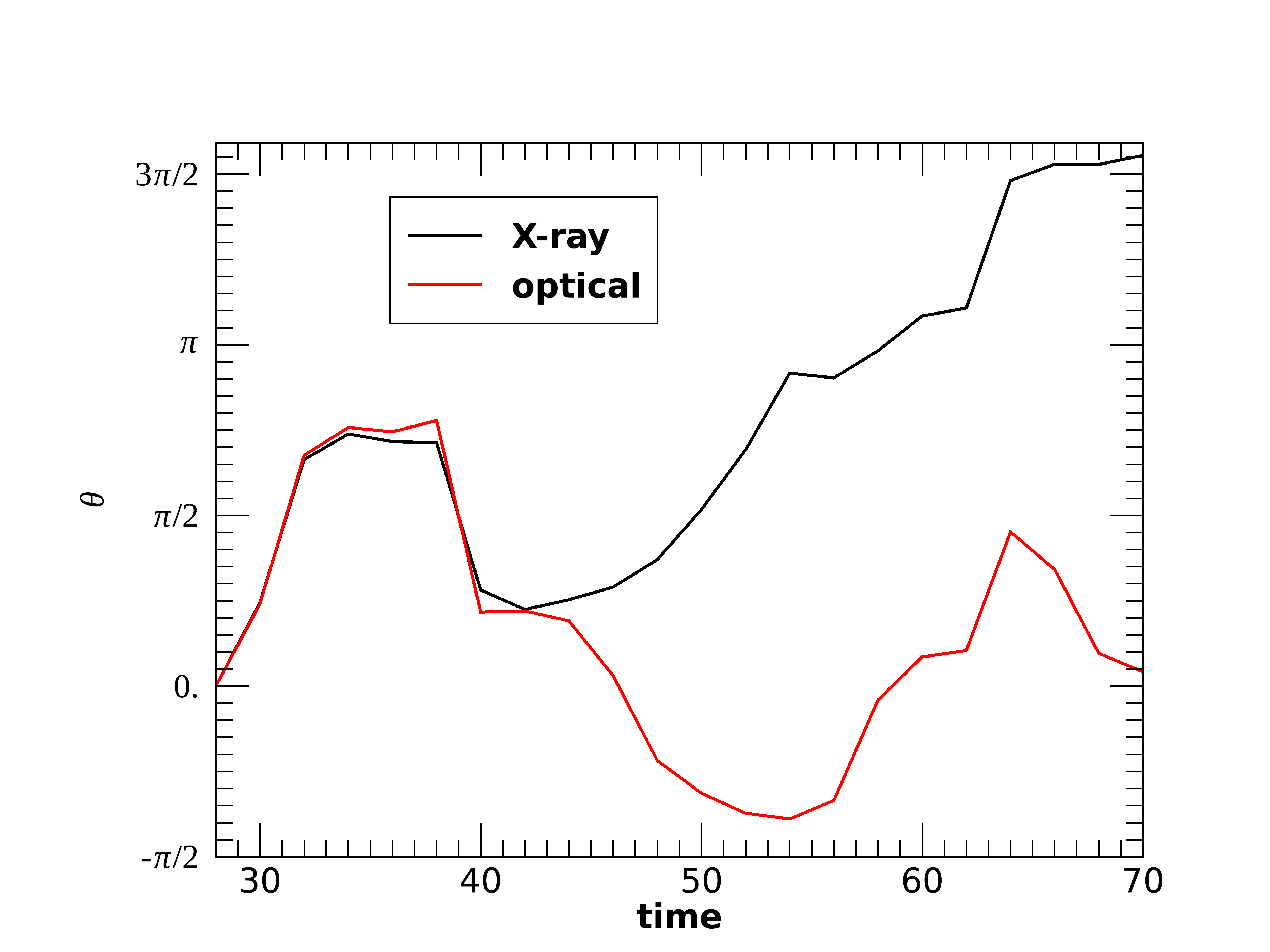} 
  \caption{Plot of the polarization angle of the integrated emission as a function of time. $\theta = 0$ corresponds to the $x-$axis direction. }
\label{fig:polangle}
  \end{figure}

In Fig \ref{fig:polangle} we plot the polarization angle as a function of time from the beginning of the flare up to $t=70$, when the emission is negligible. The polarization angle is measured counterclockwise from the positive $x$ axis. Since there is an ambiguity of $\pi$ in the definition of the angle, at each time, between values differing by $\pi$, we choose the closest to that of the previous time.  As for the polarization fraction, the optical mimics the X-rays up to $t \sim 40$,
with two swings of about $\pi/2$. Afterwards, when also the optical emission enters the decreasing phase,  they rotate in opposite directions, until they reach again the same direction at $t \sim 55$.  After that, they behave again in a similar way. As discussed above, the observed polarization  is the result of the superposition of contributions from regions having different field orientation, and the swings  in the polarization direction can be associated to a change in the prevalence of a given region.


\section{Summary and Discussion}

We have performed RMHD simulations of a magnetically dominated kink-unstable jet aimed at studying the polarimetric properties displayed by the synchrotron emission of energized electrons at different frequencies. Our study has been motivated by a previous work (T18) in which we have compared the polarization in the optical and in the X-ray band expected for particle acceleration occurring at a trans-relativistic shock (the time-dependent extension is reported in Tavecchio et al., in prep.). In that case, the properties of the system imprint a clear difference between the highly polarized X-ray emission (showing polarization fractions up to 50\%) and the much less polarized optical emission.

Our simulations show that the kink instability naturally develops current sheets at which efficient acceleration of particles into  a non-thermal distribution is expected to occur \citep{Alves18,Davelaar19}. In order to derive the properties of the radiation emitted by particles with different ages, we distinguish between freshly accelerated electrons and those injected at earlier times, which have experienced radiation losses. As in T18, the most direct application of this set-up is the modeling of BL Lac objects whose  low-energy SED component peaks in the X-ray band (the so called HBL). In this sources the emission up to the hard X-ray band is produced by relativistic electrons through synchrotron radiation, therefore carrying the imprint of the underlying magnetic field configuration.

High-energy electrons injected in the reconnection region rapidly cool due to radiation losses and therefore they are tracers of the magnetic fields in the vicinity of current sheets. Low energy and cooled particles, on the other hand, are expected to move from the original injection site following the motion of the fluid. We find that the kink instability is unable to inject strong turbulence, so the particles can only be carried away by advection, which causes only a marginal displacement from the injection regions. As a consequence, during most of the instability development, the spatial distribution of low-energy electrons does not differ much from that of younger particles. Since the fields probed by freshly, high-energy emitting particles and by the cooled ones are very similar, the properties of the polarization are predicted to be similar at all frequencies. The only difference between the two bands concerns the behavior of the angle of polarization after the maximum of the optical flare (Fig. \ref{fig:polangle}). In fact, while the angles in the X-rays and in the optical bands at earlier times are rather similar, after the optical maximum the electric vector rotates from about 0 degrees to $-\pi/2$ in the optical, while in the X-rays the angle increases, exceeding $\pi$ at late times (when, however, the emitted flux is very low).

As mentioned before, the synchrotron radiation emitted by electrons accelerated at a trans-relativistic shock is expected to display quite different polarization between the optical and the X-ray bands. The simultaneous measurements of the polarization in the X-rays and at lower frequencies (i.e. optical) therefore provides a powerful tool to test the two scenarios.
In our simulations, the duration of a flare as measured in the jet frame (corresponding to the dissipation phase) is of the order of $30$ $r_j/c$ (see Fig~\ref{fig:flux})\footnote{However, fragmentation of the current sheets into plasmoids could originate faster variability timescales.}. For a Doppler factor $D=10$ the observed duration would therefore be $t_{\rm fl}\simeq r_j/cD= 10^6 r_{j,16}$ s. Taking as a benchmark  Mkn 421, one of the brightest HBL in the X-ray band, with a flux during flares exceeding $10^{-10}$ erg cm$^{-1}$ s$^{-1}$, {\it IXPE} would be able to reach a minimum detectable polarization smaller than 20\% for exposure times as short as $10^3$ s (e.g. \citealt{Tavecchioixpe2020}). Since the average degree of polarization during this phase is around 25\% (see Fig~\ref{fig:polfrac}) and the evolution of the polarization angle occurs on a much longer timescale one can comfortably trace the evolution of the polarization fraction during the X-ray flare and compare it with the optical band, expected to closely follow the evolution displayed by the X-ray emission, for kink-induced dissipation. The close similarity between the polarization properties in the optical and in the X-rays predicted for the kink scenario is different from the evolution anticipated for the shock scenario, in which the optical emission is expected to show a lower degree of polarization (T18; Tavecchio et al. in preparation).

The relative role of the two potential candidates for dissipation, i.e. kink-induced reconnection or shocks, is thought to depend on the magnetization of the flow. High magnetization strongly favors the kink instability and reconnection. In this framework it is interesting to note that the comparison between polarimetric monitoring of blazars and the patterns expected for relativistic shocks seems to support high magnetization values \citep{Zhang16}. However, magnetized shocks are thought to be rather inefficient particle accelerators, since efficient particle diffusion is inhibited by the strong fields \citep[e.g.][]{Sironi15}. 

We also note that, contrary to what may be expected, our simulations show that appreciable turbulence develops only at quite late times, when most of the dissipation already occurred and strong current sheets are no longer present (see also \citealt{Bromberg19}). As a consequence, classical approaches in which the polarization properties are described in terms of random walk/stochastic models, which consider the contribution of several independent cells (e.g. \citealt{Jones1988,marscher2014,Kiehlmann2016,Kiehlmann2017}), are unsuitable to model the radiation produced during the development of the kink instability. A weak level of turbulence may however be present as a remnant of previous events and may somewhat lower the observed polarization degree.

Our simulation set-up is similar to that used by \cite{Zhang17}, who performed a study of the polarization properties of the emission triggered by the kink instability. We note that \cite{Zhang17} did not make a distinction between freshly accelerated and cooled particles, but only calculated the polarization considering the magnetic field properties in the regions identified as injection sites (in our treatment, this would correspond to the radiation of particles emitting at high energy). As discussed before, for the identification of the injection sites they made use of the quantity $\mathbf{J\cdot E}$, which could include not only genuine dissipation, but also the (reversible) work done by the magnetic field on the fluid.
 Another difference with respect to our study is that \cite{Zhang17} assumed {\it ad hoc} the existence of a non-thermal population of background particles, besides those energized by the dissipation process. The presence of such a population gives a quite large (around $40\%$) polarization at early times, when the field is dominated by the {\it regular} toroidal component. While the instability develops, most of the dissipation occurs around the central axis, where the magnetic field is dominated by the poloidal component, determining a rotation of the polarization angle by $90$ deg and a significant drop of the polarization fraction \citep{Zhang17}. This effect does not appear in our simulations, since at the beginning (i.e., before dissipation starts) there are no emitting particles. We finally note that, even if it were to exist, a background particle population would probably be relatively cold, therefore possibly contributing at low energies but likely not visible in the X-ray band.

Finally, we remark that, due to the limited dynamic range in the ratio of current sheet thickness to its length (the latter being of order of the jet radius), our current sheets do not develop  reconnection plasmoids. The presence of plasmoids, and associated mergers, may itself imprint interesting polarization signatures, even for an isolated reconnection layer, as described by PIC simulations \citep{zhang_18,hosking_20}.

\section*{Acknowledgments}
GB and FT acknowledge contribution from the grant INAF CTA--SKA ``Probing particle acceleration and $\gamma$-ray propagation with CTA and its precursors'' and the INAF Main Stream project ``High-energy extragalactic astrophysics: toward the Cherenkov Telescope Array''.
GB acknowledges also support from {\it{PRIN MIUR 2015}} (grant number 2015L5EE2Y). LS acknowledges support from the Sloan Fellowship, the Cottrell Fellowship, DoE DE-SC0016542, NASA ATP NNX17AG21G and NSF PHY-1903412. We acknowledge  support by the {\it{Accordo Quadro INAF-CINECA 2017}} for the availability of high performance computing resources. 

\section{Data Availability}
Data available on request

\appendix
\section{Dependence on the threshold}
In this Appendix, we discuss how the polarization degree in the X-ray band depends on our choice of the threshold on $s$ for injecting particles. Given the discussion in the main text (end of Section \ref{sec:simul}), we expect that larger values of $s$ would correspond to current sheets with a weaker level of guide field. As shown in \fig{scale}, a linear relation exists between the $s$ parameter and the local electric current. The energy injection rate of accelerated particles is assumed to scale linearly with the local dissipation rate, and so $\propto J^2$. It follows that, even though regions with large $s$ may be  rare, and occupy only a small fraction of the volume, their role in particle injection may still be appreciable.

\begin{figure}
  \includegraphics[width=\columnwidth]{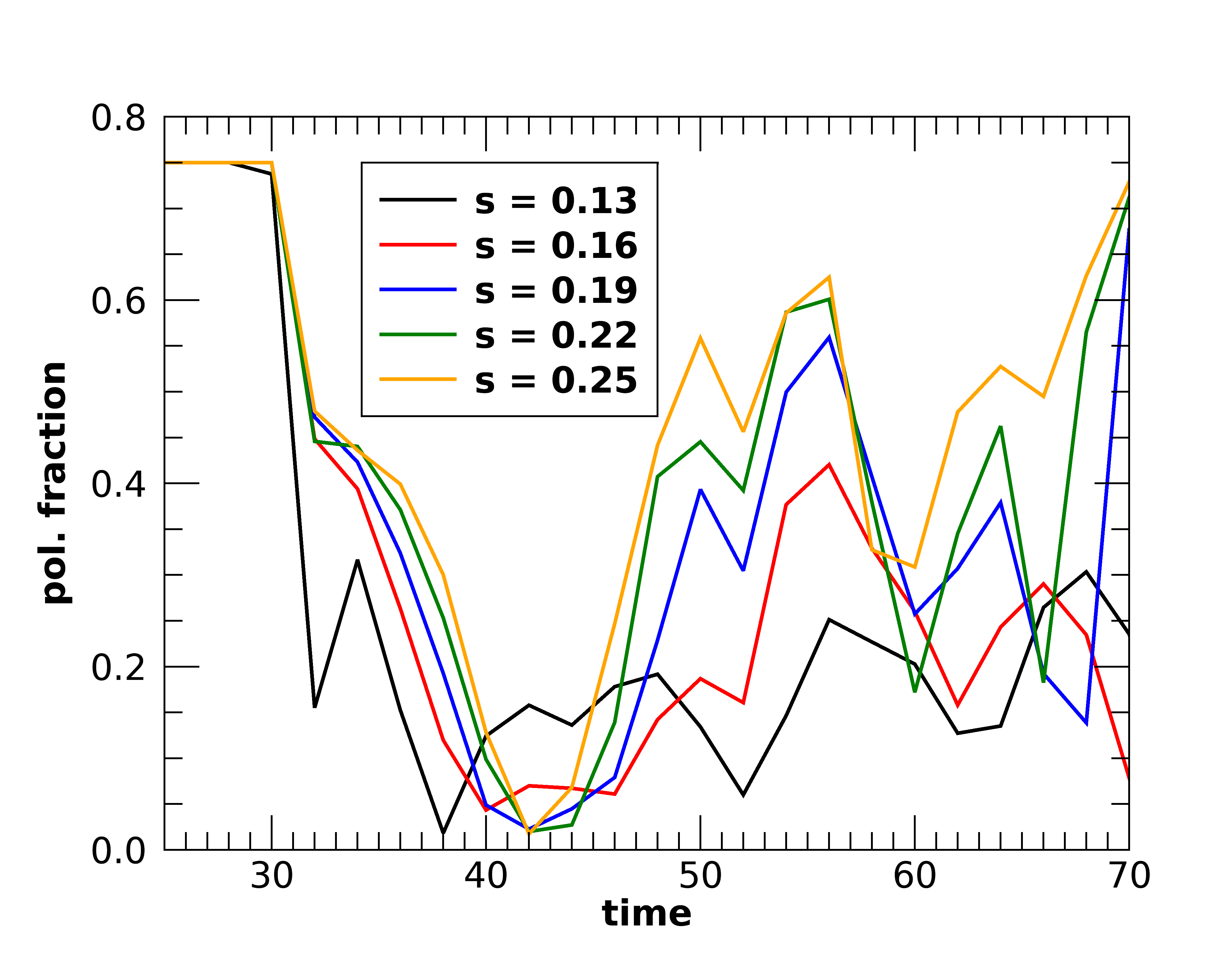} 
  \caption{Plot of the polarization degree in the X-ray band as a function of time. The different curves correspond to different values of the threshold in $s$, as indicated in the legend. }
\label{fig:polfrac1}
  \end{figure}

In \fig{polfrac1}, we compare the temporal evolution of the polarization degree among cases with different choices of the threshold in $s$, as indicated in the legend. We remind the reader that the X-ray flux starts rising at $t\sim 30$, peaks at $t\sim 35$, and is less than $0.1$ of the peak at $t\gtrsim 50$. The red to yellow curves in \fig{polfrac1} (i.e., excluding the black line) follow a continuous trend: higher thresholds in $s$ identify fewer regions suitable for particle injection; as a result, the emission is  contributed by a smaller fraction of the volume, and a larger polarization degree is attained (i.e., the polarization signal is not diluted by the simultaneous contribution of multiple regions with varying polarization angle). The black line ($s=0.13$) is still mostly consistent with this trend (so, it is typically below the other curves), aside from the time range $40\lesssim t \lesssim 50$. We speculate that, at these times, most of the emission might be contributed by a single region with $s\sim 0.13$, which controls the overall polarized flux.

\end{document}